\documentclass[sigconf, noacm, dvipsnames]{acmart}

\AtBeginDocument{%
  \providecommand\BibTeX{{%
    \normalfont B\kern-0.5em{\scshape i\kern-0.25em b}\kern-0.8em\TeX}}}

\newcommand\ignore[1]{ }

\usepackage{xcolor}
\usepackage{multirow}
\usepackage[colorinlistoftodos,prependcaption,textsize=small]{todonotes}

\definecolor{airforceblue}{rgb}{0.36, 0.54, 0.66}
\definecolor{dodgerblue}{rgb}{0.12, 0.56, 1.0}
\definecolor{brandeisblue}{rgb}{0.0, 0.44, 1.0}
\definecolor{brickred}{rgb}{0.8, 0.25, 0.33}
\definecolor{eggplant}{rgb}{0.38, 0.25, 0.32}
\definecolor{byzantium}{rgb}{0.44, 0.16, 0.39}
\newcommand{\jgl}[1]{\textcolor{brickred}{\textit{JGL: #1}}}
\newcommand{\sylvan}[1]{\textcolor{eggplant}{#1}}

\newcommand{\julien}[1]{\textcolor{RedOrange}{#1}}

\definecolor{dgreen}{rgb}{0.00, 0.75, 0.00}
\definecolor{ddgreen}{rgb}{0.00, 0.50, 0.00}

\newcommand{\juan}[1]{\textcolor{black}{#1}}
\newcommand{\juang}[1]{\textcolor{black}{#1}}
\newcommand{\juangg}[1]{\textcolor{black}{#1}}
\newcommand{\juanggr}[1]{\textcolor{black}{#1}}

\usepackage{tikz}
\usetikzlibrary{shapes.geometric, arrows, positioning, shadows}

\pgfdeclarelayer{background}
\pgfsetlayers{background,main}

\newcommand{\vspacing}{0.8cm}

\tikzstyle{texte} = [above]
\newcommand{\background}[5]{%
  \begin{pgfonlayer}{background}
    \path (#1.west |- #2.north)+(-0.5,0.5) node (a1) {};
    \path (#3.east |- #4.south)+(+0.5,-0.25) node (a2) {};
    \path[fill=yellow!20,rounded corners, draw=black!50, dashed]
      (a1) rectangle (a2);
    \path (a1.east |- a1.south)+(1.2,-0.3) node (u1)[texte]
      {\scriptsize\textit{DPU execution}};
  \end{pgfonlayer}}

\tikzstyle{stop} = [rectangle, rounded corners, minimum width=3cm, minimum height=1cm,text centered, draw=black, fill=violet!30, drop shadow]
\tikzstyle{io} = [trapezium, trapezium left angle=70, trapezium right angle=110, text centered, text width=5cm, draw=black, fill=blue!30, drop shadow]
\tikzstyle{processcpu} = [rectangle, text centered, text width=5cm, draw=black, fill=orange!30, drop shadow]
\tikzstyle{processdpu} = [rectangle, text centered, text width=5cm, draw=black, fill=red!30, drop shadow]
\tikzstyle{decision} = [diamond, aspect=2, text centered, draw=black, fill=green!30, drop shadow]
\tikzstyle{arrow} = [thick,->,>=stealth]

\usepackage{algorithm}
\usepackage{algpseudocode}
\usepackage{stmaryrd}
\usepackage{amsmath}
\usepackage{listings}
\usepackage{subcaption}

\definecolor{mygreen}{rgb}{0,0.6,0}
\definecolor{mygray}{rgb}{0.5,0.5,0.5}
\definecolor{mymauve}{rgb}{0.58,0,0.82}

\lstdefinestyle{myC}{
  language=Matlab,
  backgroundcolor=\color{white},  
  basicstyle=\footnotesize,        
  breakatwhitespace=false,  
  breaklines=true,       
  captionpos=b,                   
  commentstyle=\color{mygreen},    
  deletekeywords={...},           
  escapechar=\%,
  xleftmargin=0pt,
  xrightmargin=0pt,
  aboveskip=\medskipamount,
  belowskip=\medskipamount,
  extendedchars=true,            
  keepspaces=true,               
  keywordstyle=\color{blue},      
  language=C++,              
  morekeywords={__builtin_mul_sl_ul_rrr,__builtin_mul_sl_sh_rrr,mul_ul_ul,mul_sh_ul,mul_sh_sh,mul_sl_ul,mul_sl_sh,lbs,lhs,move,lsl_add,lw,add,sw,jneq,mem_alloc, *,...},          
  numbers=left,                   
  numbersep=1pt,                   
  numberstyle=\tiny\color{mygray}, 
  rulecolor=\color{black},     
  showspaces=false,                
  showstringspaces=false,          
  showtabs=false,                  
  stepnumber=1,                    
  stringstyle=\color{mymauve},     
  tabsize=2,	                   
  title=\lstname                   
}

\definecolor{bluehl}{rgb}{0.8,0.874,1}
\definecolor{pinkhl}{rgb}{0.992156863,0.847058824,1}
\definecolor{macaroniandcheese}{rgb}{1.0, 0.74, 0.53}
\definecolor{mossgreen}{rgb}{0.68, 0.87, 0.68}
\definecolor{greenhl}{rgb}{0.835,0.996,0.939}
\definecolor{yellowhl}{rgb}{0.996,0.957,0.8}
\definecolor{palecerulean}{rgb}{0.61, 0.77, 0.89}
\definecolor{gray(x11gray)}{rgb}{0.75, 0.75, 0.75}

\newcommand{\HilightBlue}{\makebox[0pt][l]{\color{palecerulean}\rule[-4pt]{\linewidth}{10pt}}}
\newcommand{\HilightPink}{\makebox[0pt][l]{\color{macaroniandcheese}\rule[-4pt]{\linewidth}{10pt}}}
\newcommand{\HilightYellow}{\makebox[0pt][l]{\color{yellowhl}\rule[-4pt]{\linewidth}{10pt}}}
\newcommand{\HilightGreen}{\makebox[0pt][l]{\color{mossgreen}\rule[-4pt]{\linewidth}{10pt}}}

\newcommand{\HilightGray}{\makebox[0pt][l]{\color{gray(x11gray)}\rule[-4pt]{\linewidth}{10pt}}}

\AtBeginDocument{\DeclareCaptionSubType{lstlisting}}

\setcopyright{none}
\renewcommand\footnotetextcopyrightpermission[1]{}

\makeatletter
\def\bstctlcite{\@ifnextchar[{\@bstctlcite}{\@bstctlcite[@auxout]}}
\def\@bstctlcite[#1]#2{\@bsphack
  \@for\@citeb:=#2\do{%
    \edef\@citeb{\expandafter\@firstofone\@citeb}%
    \if@filesw\immediate\write\csname #1\endcsname{\string\citation{\@citeb}}\fi}%
  \@esphack}
\makeatother

\usepackage{xspace}
\newcommand{\floatp}{floating-point\xspace}
\newcommand{\fixedp}{fixed-point\xspace}
\newcommand{\km}{K-Means\xspace}

\ignore{
\newcommandx{\unsure}[2][1=]{\todo[linecolor=red,backgroundcolor=red!25,bordercolor=red,#1, size=\tiny]{#2}}
\newcommandx{\feedback}[2][1=]{\todo[linecolor=yellow,backgroundcolor=yellow!25,bordercolor=yellow,#1]{#2}}

\titlespacing\section{2pt}{3pt plus 1pt minus 1pt}{2pt plus 1pt minus 1pt}
\titlespacing\subsection{2pt}{3pt plus 1pt minus 1pt}{2pt plus 1pt minus 1pt}
\titlespacing\subsubsection{2pt}{3pt plus 1pt minus 1pt}{2pt plus 1pt minus 1pt}

\makeatletter
\g@addto@macro{\normalsize}{%
  \setlength{\abovedisplayskip}{2pt plus 1pt minus 1pt}
  \setlength{\belowdisplayskip}{2pt plus 1pt minus 1pt}
  \setlength{\abovedisplayshortskip}{0pt}
  \setlength{\belowdisplayshortskip}{0pt}
  \setlength{\intextsep}{2pt plus 1pt minus 1pt}
  \setlength{\textfloatsep}{3pt plus 1pt minus 1pt}
  \setlength{\dbltextfloatsep}{3pt plus 1pt minus 1pt}
  \setlength{\skip\footins}{4pt plus 1pt minus 1pt}}
  \setlength{\abovecaptionskip}{2pt plus 1pt minus 1pt}
\makeatother
}

\newcommand{\tsc}[1]{\textsuperscript{#1}} 
\newcommand{\affilETH}{\tsc{1}}
\newcommand{\affilUPM}{\tsc{2}}

\begin{document}
\bstctlcite{IEEEexample:BSTcontrol}

\title{An Experimental Evaluation of Machine Learning Training \\ on a Real Processing-in-Memory System}

\author{
 {%
     Juan Gómez-Luna$^1$\quad 
     Yuxin Guo$^1$\quad 
     Sylvan Brocard$^2$\quad 
     Julien Legriel$^2$
 }
}
\author{
 {
     Remy Cimadomo$^2$\quad
     Geraldo F. Oliveira$^1$\quad
     Gagandeep Singh$^1$\quad
     Onur Mutlu$^1$
 }
}


\affiliation{
\institution{
      \vspace{5pt}
      \affilETH ETH Zürich \quad
      \affilUPM UPMEM \quad
  }
}

\pagestyle{plain}

\begin{abstract}

Training machine learning algorithms is a computationally intensive process, which is frequently memory-bound due to repeatedly accessing large training datasets. 
As a result, processor-centric systems (e.g., CPU, GPU) suffer from costly data movement between memory units and processing units, which consumes large amounts of energy and execution cycles. 
Memory-centric computing systems, i.e., computing systems with processing-in-memory (PIM) capabilities, can alleviate this data movement bottleneck.

Our goal is to understand the potential of modern general-purpose PIM architectures to accelerate machine learning training. 
To do so, we (1) implement several representative classic machine learning algorithms (namely, linear regression, logistic regression, decision tree, {\km} clustering) on a real-world general-purpose PIM architecture, (2) {rigorously evaluate and} characterize them in terms of accuracy, performance and scaling, and (3) compare to their counterpart implementations on CPU and GPU. 
Our experimental evaluation on a {real} memory-centric computing system with more than 2500 PIM cores shows that general-purpose PIM architectures can greatly accelerate memory-bound machine learning workloads, when the necessary operations and datatypes are natively supported by PIM hardware. 
For example, our PIM implementation of decision tree is \juangg{between $27\times$ and $113\times$} faster than a state-of-the-art CPU version on an 8-core Intel Xeon, and \juangg{between $1.34\times$ and $4.5\times$} faster than a state-of-the-art GPU version on an NVIDIA A100. 
Our \juan{PIM implementation of} {\km} clustering is $2.8\times$ and $3.2\times$ \juan{faster} than CPU and GPU implementations, respectively.

To our knowledge, our work is the first one to evaluate training of machine learning algorithms on a real-world general-purpose PIM architecture. 
We conclude this paper with several key observations, takeaways, and recommendations that can inspire users of machine learning workloads, programmers of PIM architectures, and hardware designers and architects of future memory-centric computing systems. 
\juan{We open-source all our code and datasets at~\url{https://github.com/CMU-SAFARI/pim-ml}.}

\end{abstract}



\maketitle

\section{Introduction}

Machine learning (ML) algorithms~\cite{geron2019, alpaydin2020, goodfellow2016, mohri2018, shalev2014, raschka2019} have become ubiquitous in many fields of science and technology due to their ability to learn {from} and 
{improve with} experience with minimal human intervention. These algorithms train by updating their model parameters in an iterative manner to improve the overall prediction accuracy.  
However, training ML algorithms is a computationally intensive process, which requires large amounts of training data~\cite{deoliveira2021IEEE,wang2020survey, dunner2018snap}. 
Accessing training data in current processor-centric systems (e.g., CPU, GPU) \juan{requires} costly data movement between memory and processors, which results in high energy consumption and a large percentage of the total execution cycles. 
This data movement can become the bottleneck of the training process, if there is not enough computation and locality to amortize its cost~\cite{xie2017cumf_sgd, de2017understanding, kim2021gradpim, mahajan2016tabla, bo2019cluster, bender2015memsys}. 

One way to alleviate the cost of data movement is \emph{processing-in-memory} (\emph{PIM})~\cite{mutlu2019,mutlu2020modern, ghoseibm2019, seshadri2020indram, mutlu2019enabling}, a {data-centric} computing paradigm that places processing elements near or inside the memory arrays. 
PIM has been explored for decades~\cite{stone1970logic, Kautz1969, shaw1981non, kogge1994, gokhale1995processing, patterson1997case, oskin1998active, kang1999flexram, Mai:2000:SMM:339647.339673,murphy2001characterization, Draper:2002:ADP:514191.514197,aga.hpca17,eckert2018neural,fujiki2019duality,kang.icassp14,seshadri.micro17,seshadri.arxiv16,Seshadri:2015:ANDOR,seshadri2013rowclone,angizi2019graphide,kim.hpca18,kim.hpca19,gao2020computedram,chang.hpca16,xin2020elp2im,li.micro17,deng.dac2018,hajinazarsimdram,rezaei2020nom,wang2020figaro,ali2019memory,li.dac16,angizi2018pima,angizi2018cmp,angizi2019dna,levy.microelec14,kvatinsky.tcasii14,shafiee2016isaac,kvatinsky.iccd11,kvatinsky.tvlsi14,gaillardon2016plim,bhattacharjee2017revamp,hamdioui2015memristor,xie2015fast,hamdioui2017myth,yu2018memristive,syncron,fernandez2020natsa,cali2020genasm,kim.bmc18,ahn.pei.isca15,ahn.tesseract.isca15,boroumand.asplos18,boroumand2019conda,singh2019napel,asghari-moghaddam.micro16,DBLP:conf/sigmod/BabarinsaI15,chi2016prime,farmahini2015nda,gao.pact15,DBLP:conf/hpca/GaoK16,gu.isca16,guo2014wondp,hashemi.isca16,cont-runahead,hsieh.isca16,kim.isca16,kim.sc17,DBLP:conf/IEEEpact/LeeSK15,liu-spaa17,morad.taco15,nai2017graphpim,pattnaik.pact16,pugsley2014ndc,zhang.hpdc14,zhu2013accelerating,DBLP:conf/isca/AkinFH15,gao2017tetris,drumond2017mondrian,dai2018graphh,zhang2018graphp,huang2020heterogeneous,zhuo2019graphq,santos2017operand,ghoseibm2019,wen2017rebooting,besta2021sisa,ferreira2021pluto,olgun2021quactrng,lloyd2015memory,elliott1999computational,zheng2016tcam,landgraf2021combining,rodrigues2016scattergather,lloyd2018dse,lloyd2017keyvalue,gokhale2015rearr,nair2015active,jacob2016compiling,sura2015data,nair2015evolution,balasubramonian2014near,xi2020memory,impica,boroumand2016pim,giannoula2022sparsep,giannoula2022sigmetrics,denzler2021casper,boroumand2021polynesia,boroumand2021icde,singh2021fpga,singh2021accelerating,herruzo2021enabling,yavits2021giraf,asgarifafnir,boroumand2021google_arxiv,boroumand2021google,amiraliphd,singh2020nero,seshadri.bookchapter17,diab2022high,diab2022hicomb,fujiki2018memory,zha2020hyper,mutlu.imw13,mutlu.superfri15,ahmed2019compiler,jain2018computing,ghiasi2022genstore,deoliveira2021IEEE,deoliveira2021,cho2020mcdram,shin2018mcdram,gu2020ipim,lavenier2020,Zois2018}. However, memory technology challenges prevented from its successful materialization in commercial products. For example, the limited number of metal layers in DRAM~\cite{weber2005current,peng2015design} makes conventional processor designs 
{impractical} {in commodity DRAM chips}~\cite{devaux2019,yuffe2011,christy2020,singh2017}.

{Real-world} PIM systems have 
only recently been manufactured~\cite{upmem,upmem2018,gomezluna2021benchmarking, gomezluna2022ieeeaccess, gomezluna2021cut, kwon202125, lee2021hardware, ke2021near, lee2022isscc, niu2022isscc}. 
The UPMEM company, for example, 
{introduced} the first general-purpose {commercial} PIM architecture~\cite{upmem,upmem2018,gomezluna2021benchmarking, gomezluna2022ieeeaccess, gomezluna2021cut}, which integrates small in-order cores near DRAM memory banks. 
High-bandwidth memory (HBM)-based HBM-PIM~\cite{kwon202125, lee2021hardware} and Acceleration DIMM (AxDIMM)~\cite{ke2021near} are Samsung's proposals that have been successfully tested 
{via real} prototypes. HBM-PIM features \emph{Single Instruction Multiple Data} (\emph{SIMD}) units, which 
{support} multiply-add and multiply-accumulate operations, near the banks in HBM layers~\cite{jedec.hbm.spec, lee.taco16}, and it is designed to accelerate neural network inference. AxDIMM is a near-rank solution that places an FPGA fabric on {a DDR module} 
to accelerate specific workloads (e.g., recommendation inference). 
Accelerator-in-Memory (AiM)~\cite{lee2022isscc} is a GDDR6-based PIM architecture from SK Hynix with specialized units for multiply-accumulate and activation functions for deep learning. 
HB-PNM~\cite{niu2022isscc} is a 3D-stacked-based PIM architecture from Alibaba, which stacks a layer of LPDDR4 memory and a logic layer with specialized accelerators for {recommendation} systems.

Our \emph{goal} in this work is to quantify the potential of general-purpose PIM architectures for training of machine learning algorithms. 
To this end, we implement four representative {classical} machine learning algorithms (linear regression~\cite{freedman2009statistical, yan2009linear}, logistic regression~\cite{freedman2009statistical, hosmer2013applied}, decision tree~\cite{suthaharan2016decision}, \km clustering~\cite{Lloyd82leastsquares}) 
{on} a general-purpose memory-centric system containing PIM-enabled memory, 
{specifically} the UPMEM PIM architecture~\cite{upmem,upmem2018,gomezluna2021benchmarking, gomezluna2022ieeeaccess, gomezluna2021cut}. 
We do \emph{not} include training of deep learning 
algorithms in our study, since GPUs {and TPUs} have a solid position as the preferred {and highly optimized accelerators for 
deep learning} training~\cite{hwukirk2016.nn,chetlur2014cudnn,gao2017tetris,abadi2016tensorflow,runai_gpu,jouppi2017datacenter, jouppi2021ten} 
\juan{due to their extremely high floating-point performance.}\footnote{\juan{The UPMEM PIM architecture (used in this study) currently does not have native support for floating-point operations~\cite{devaux2019,upmem,upmem2018,gomezluna2021benchmarking, gomezluna2022ieeeaccess, gomezluna2021cut}.}}

Our PIM implementations of ML algorithms follow {PIM} programming recommendations in recent literature~\cite{upmem-guide, upmem2018, gomezluna2021benchmarking, gomezluna2022ieeeaccess}. 
We apply several optimizations to overcome the limitations of existing general-purpose PIM architectures (e.g., limited instruction set, relatively simple pipeline, relatively low frequency) and take full advantage of the inherent strengths of PIM (e.g., large memory bandwidth, {low memory latency}). 

We evaluate our PIM implementations in terms of {training} accuracy, performance, and scaling characteristics on a {real} memory-centric system with PIM-enabled memory~\cite{upmem, upmem-sdk, upmem-guide}. 
We run our experiments on a real-world PIM system~\cite{upmem} with 2,524 PIM cores running at 425 MHz, and 158 GB of DRAM memory.

Our experimental {real system} evaluation provides 
{new} observations and insights, 
{including the following:} 
\begin{itemize}
\item ML training workloads that 
{show memory-bound behavior} in processor-centric systems can greatly benefit from (1) \fixedp {data} representation, (2) quantization~\cite{zmora2021quantization,gholamisurvey}, and (3) hybrid precision implementation~\cite{hopper, lee2022isscc} {(without much accuracy loss) in PIM systems}, in order to 
{alleviate} the lack of native support for \floatp 
and high-precision (i.e., 32- and 64-bit) {arithmetic operations} {in the evaluated PIM system}. 
\item ML training workloads that require complex activation functions (e.g., sigmoid)~\cite{han1995influence} can take advantage of \emph{lookup tables} (\emph{LUTs})~\cite{ferreira2021pluto, deng2019lacc, gao2016draf} {in PIM systems} instead of function approximation (e.g., Taylor series)~\cite{weisstein2004taylor}, when 
{PIM systems} lack native support for those activation functions. 
\item 
{Data can be placed and laid out such that} accesses of PIM cores to their nearby memory banks are 
streaming, 
{which enables} better exploitation of the \juan{internal} PIM memory bandwidth. 
\item ML training workloads with large training datasets can greatly benefit from scaling {the size of} PIM-enabled memory with PIM cores attached to memory banks. Training datasets can remain in memory without being moved to the host processor (e.g., CPU, GPU) in every iteration of the training process. Even if PIM cores need to communicate intermediate results via the host processor, this communication overhead is 
tolerable. 
\end{itemize}



We compare our PIM implementations of linear regression, {logistic} regression, decision tree, and \km clustering to their state-of-the-art CPU and GPU counterparts. We observe that memory-centric systems with PIM-enabled memory can significantly outperform processor-centric systems for memory-bound ML training workloads, when the operations needed by the ML workloads are natively supported by PIM hardware (or can be replaced by {efficient LUT implementations}). 

We {aim to} open-source all our PIM implementations of ML training workloads, training datasets, and evaluation scripts \juan{in our GitHub repository~\cite{gomezluna2023repo}}.

\section{Background}
\label{sec:background}
In this section, we first present a general classification of machine learning workloads, and the motivation for accelerating them in processing-in-memory (PIM) systems. 
Second, we give an introduction to 
PIM, and the current real-world PIM systems (as of \juan{April 2023}) with a focus on the UPMEM PIM architecture, which we employ in our study.

\subsection{Machine Learning Workloads}
\label{sec:machinelearning}
Machine learning (ML)~\cite{geron2019, alpaydin2020, goodfellow2016, mohri2018, shalev2014, raschka2019} is a family of algorithms that 
learns a target function (or \emph{model}) that best maps the input variables to an output variable. 
ML algorithms build (\emph{train}) a model using the observed data (\emph{training dataset}). The model is then used to make (\emph{infer}) predictions or decisions. 

\begin{figure}[h]
\centering
\includegraphics[width=1.0\linewidth]{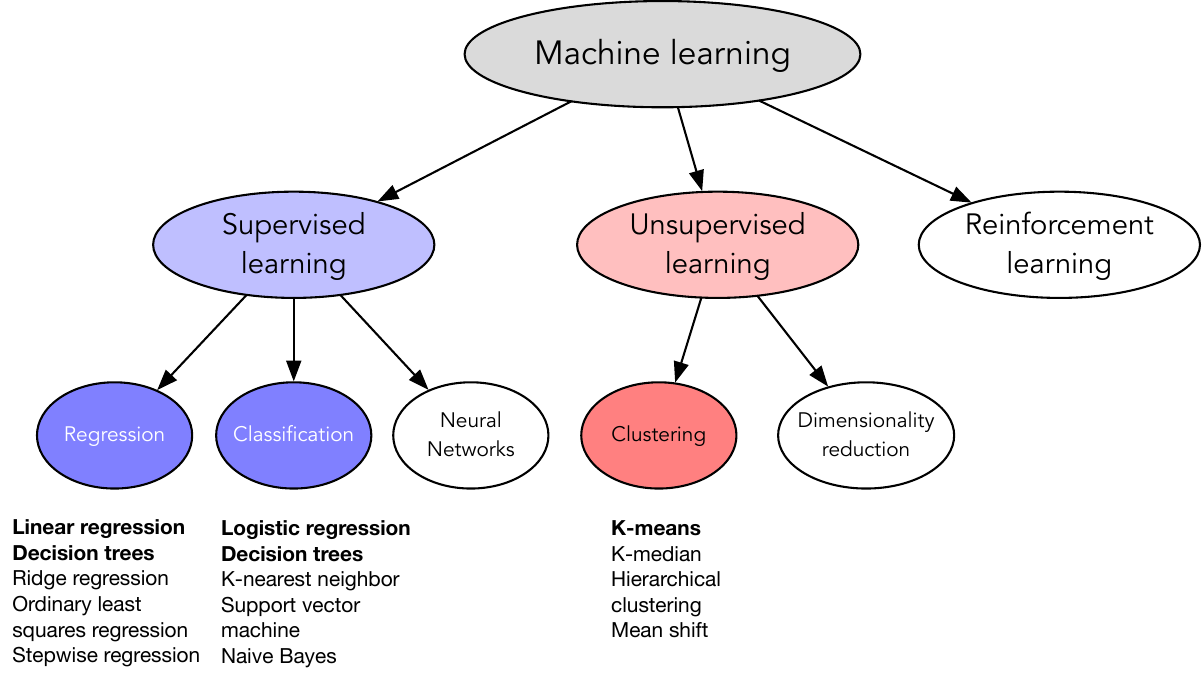}
\caption{Machine learning algorithms. Colored categories (specifically, bold-font algorithms) are the focus of this work.}
\label{fig:mlmap}
\end{figure}

Machine learning is commonly divided into three main categories of algorithms: (1) \emph{supervised learning}, (2) \emph{unsupervised learning}, and (3) \emph{reinforcement learning}, as shown in Figure~\ref{fig:mlmap}. 
Supervised algorithms train a model using 
training datasets that contain input \emph{features} with expected labeled 
outputs. We divide supervised learning into \emph{classification}, \emph{regression}, and \emph{neural networks}. 
Unsupervised algorithms learn to find structure or commonalities in the data (e.g., grouping, clustering) without using labeled or classified data. We divide unsupervised learning into \emph{clustering} and \emph{dimensionality reduction}. 
Reinforcement learning algorithms train an \emph{agent} to achieve an objective by interacting with its \emph{environment}. Thus, they gather training data from this interaction. 

ML training with large amounts of data is a computationally expensive process~\cite{deoliveira2021IEEE,wang2020survey, dunner2018snap}. 
This process requires several iterations to update an ML model's parameters.  Therefore, hardware accelerators (e.g., general-purpose GPU~\cite{hwukirk2016.nn}, domain-specific accelerators~\cite{jouppi2017datacenter, abts2020think, rocki2020fast}) are widely used to speed up training. 
Recent research~\cite{kim2016isca, liu2018processing} proposes 
PIM approaches to deal with the frequent data movement between memory and processing elements (either general-purpose cores or accelerators) that is needed to access training data. 
PIM approaches are especially effective for ML algorithms (or parts of an ML algorithm) where the amount of computation is not enough to amortize the cost of moving training data to the processing elements. Such workloads typically have (1) low \emph{arithmetic intensity} (i.e., arithmetic instructions executed per byte accessed from memory), (2) low temporal locality, and/or (3) irregular memory accesses (e.g., due to sparsity of data)~\cite{deoliveira2021IEEE}. 
However, PIM-based accelerator proposals are usually tailored to specific ML algorithms (e.g., CNNs~\cite{kim2016isca, liu2018processing}). As a result, they are less efficient 
for other types of ML workloads.

Our \textbf{goal} in this study is to analyze how real-world general-purpose PIM architectures can accelerate training of representative ML algorithms, and generate insights and recommendations that are useful to programmers and architecture designers. 
We select four representative 
classic machine learning algorithms (linear regression, logistic regression, decision tree, \km clustering) from three of the subcategories (regression, classification, clustering) in Figure~\ref{fig:mlmap}. 

These four workloads have been shown to be memory-bound in processor-centric systems (e.g., CPU, GPU) due to their low arithmetic intensity and low data reuse. 
Linear regression and logarithmic regression make use of \emph{gradient descent}~\cite{polyak1987} or \emph{stochastic gradient descent}~\cite{boyd2004convex} as optimization algorithms during training. Recent literature~\cite{xie2017cumf_sgd, de2017understanding, kim2021gradpim, mahajan2016tabla} shows that both are memory-bound. 
Decision tree and \km clustering are also memory-bound~\cite{bo2019cluster, bender2015memsys} due to their frequent memory accesses and lightweight computation (mainly comparisons in decision tree, or a few additions and multiplications in \km). 

\begin{figure}[ht]
    \centering
    \includegraphics[width=1.0\linewidth]{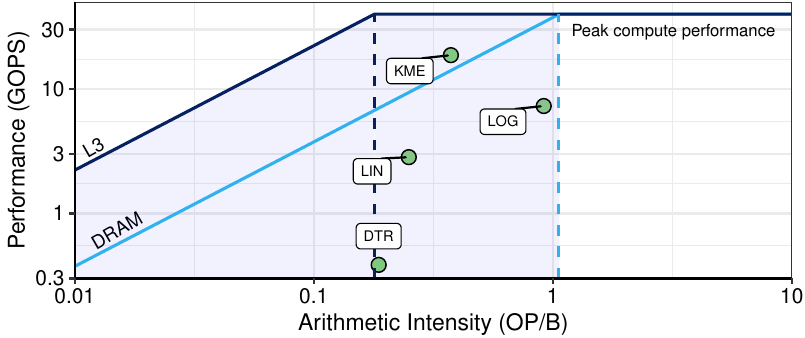}
    \vspace{-20pt}
    \caption{Roofline model for the CPU versions of four ML workloads \juan{(LIN: linear regression, LOG: logistic regression, DTR: decision tree, KME: {\km} clustering)} on an Intel Xeon E3-1225 v6 CPU.}
    \label{fig:roofline}
\end{figure}

We employ the roofline model~\cite{roofline} to quantify the memory boundedness of the CPU versions of the four workloads. Figure~\ref{fig:roofline} shows the roofline model on an Intel Xeon E3-1225 v6 CPU~\cite{xeon-e3-1225} with Intel Advisor~\cite{advisor}. We observe from Figure~\ref{fig:roofline} that all of the CPU versions of the four workloads are in the memory-bound area of the roofline model (i.e., the shaded region on the left side of the intersection between the DRAM bandwidth roof and the peak compute performance roof). Hence, we confirm that the four workloads are limited by memory access. As a result, these ML workloads are potentially suitable for PIM. 

We implement various versions of the four algorithms (with different datatypes and optimizations) on the first commercially-available PIM architecture, the UPMEM PIM architecture (Section~\ref{sec:pim}). We describe our implementations in Section~\ref{sec:implementation} and evaluate them in Section~\ref{sec:evaluation}.

We do \emph{not} include any neural network (or deep learning algorithm) or reinforcement learning algorithm in our study 
\juan{for two main reasons}. 
First, training of neural networks (e.g., CNN, RNN, GAN) can generally benefit from large caches and register files in processor-centric computing systems, since they expose high temporal locality~\cite{deoliveira2021IEEE}. Together with their inherent data-level parallelism \juan{and very high floating-point operation intensity}, they are a good fit for GPUs~\cite{chetlur2014cudnn}. In fact, the state-of-the-art ML-targeted PIM architecture~\cite{lee2021hardware, kwon202125} shows performance improvements for neural network inference (not training) and with small batch sizes. 
Second, reinforcement learning (RL)~\cite{sutton_2018} is an inherently sequential process, where an \emph{agent} learns to make decisions by receiving a \emph{reward} at timestep $t+1$ for an \emph{action} that was performed at timestep $t$ on an \emph{environment}. 
\juan{As a result, RL does not appear as a natural fit for PIM systems with \juang{many} parallel processing elements, such as the one depicted in Figure~\ref{fig:scheme}.} 
For deep reinforcement learning (DRL), the state-of-the-art approaches~\cite{cho2019fa3c} accelerate only the neural network training part in RL \juan{(neural network training is out of the scope of our work as explained above)}. 
However, there are three limitations of such proposals~\cite{singh2022sibyl}: (1) a major part in reinforcement learning is spent on sequential interaction between the agent and the environment, while also collecting the training dataset for the neural network, (2) the network sizes used in reinforcement learning are usually small and the weights perfectly fit in on-chip caches of the CPU, and (3) training in DRL can be done asynchronously in the background on the host CPU, while the agent is sequentially interacting with its environment.

\subsection{Processing-in-Memory}
\label{sec:pim}
Processing-in-memory (PIM) is a computing paradigm that advocates for memory-centric computing systems, where processing elements (general-purpose cores and/or accelerators) are placed near or inside the memory arrays. 
Firstly proposed more than 50 years ago~\cite{Kautz1969,stone1970logic}, processing-in-memory is a feasible solution to alleviate the \emph{data movement bottleneck}~\cite{mutlu2019,mutlu2020modern}, caused by 
(1) the need for moving data between memory units and compute units in processor-centric systems, \juan{which causes a huge performance loss and energy waste}, and worsened by (2) the \juan{increasing performance disparity between fast processor units and slow memory units}.

Recent innovations in memory technology (e.g., 3D-stacked memories~\cite{hmc.spec.2.0,jedec.hbm.spec}, nonvolatile memories~\cite{lee-isca2009, kultursay.ispass13, strukov.nature08, wong.procieee12,girard2020survey}) represent an opportunity to redesign the memory subsystem and equip it with compute capabilities. Recent research proposes processing-in-memory solutions that can be grouped into two main trends. 
\emph{Processing-near-memory} (\emph{PNM}) places processing elements near the memory arrays (e.g., in the logic layer of 3D-stacked memories or in the same chip as memory banks in 2D memories). Processing elements can be general-purpose cores~\cite{deoliveira2021,boroumand.asplos18,boroumand2019conda,ahn.tesseract.isca15,syncron,singh2019napel}, application-specific accelerators~\cite{zhu2013accelerating, DBLP:conf/isca/AkinFH15, DBLP:conf/sigmod/BabarinsaI15, kim.bmc18, cali2020genasm,fernandez2020natsa,impica,singh2020nero}, simple functional units~\cite{ahn.pei.isca15,nai2017graphpim,hadidi2017cairo}, GPU cores~\cite{zhang.hpdc14, pattnaik.pact16, hsieh.isca16,kim.sc17}, or reconfigurable logic~\cite{DBLP:conf/hpca/GaoK16, guo2014wondp, asghari-moghaddam.micro16}. 
\emph{Processing-using-memory} (\emph{PUM}) leverages the analog operational principles of memory cells in SRAM~\cite{aga.hpca17,eckert2018neural,fujiki2019duality,kang.icassp14}, DRAM~\cite{seshadri2020indram,seshadri.micro17,seshadri2013rowclone,kim.hpca18,kim.hpca19,gao2020computedram,chang.hpca16,li.micro17,hajinazarsimdram,wang2020figaro,olgun2021quactrng}, or nonvolatile memory~\cite{li.dac16,angizi2018pima,angizi2018cmp,angizi2019dna,levy.microelec14,kvatinsky.tcasii14,shafiee2016isaac,kvatinsky.iccd11,kvatinsky.tvlsi14,gaillardon2016plim,bhattacharjee2017revamp,hamdioui2015memristor,xie2015fast,hamdioui2017myth,yu2018memristive,puma-asplos2019, ankit2020panther,chi2016prime,ambrosi2018hardware,bruel2017generalize,huang2021mixed,zheng2016tcam,xi2020memory}. 
PUM enables different operations such as data copy and initialization~\cite{chang.hpca16,seshadri2013rowclone,wang2020figaro}, 
bulk bitwise operations~\cite{seshadri.micro17,li.dac16,angizi2018pima,angizi2018cmp,aga.hpca17,gao2020computedram}, 
and arithmetic operations~\cite{levy.microelec14,kvatinsky.tcasii14,aga.hpca17,li.micro17,shafiee2016isaac,eckert2018neural,fujiki2019duality,kvatinsky.iccd11,kvatinsky.tvlsi14,gaillardon2016plim,hajinazarsimdram}. 

Real-world processing-in-memory architectures are finally becoming a reality, with the commercialization of the UPMEM PIM architecture~\cite{upmem,upmem2018,gomezluna2021benchmarking}, and the announcement of \juan{Samsung} HBM-PIM~\cite{kwon202125,lee2021hardware}, \juan{Samsung} AxDIMM~\cite{ke2021near}, \juan{SK Hynix} AiM~\cite{lee2022isscc}, and \juan{Alibaba} HB-PNM~\cite{niu2022isscc} (all four prototyped and evaluated in real systems). 
The five of them are PNM architectures. 
UPMEM places small general-purpose in-order cores (called \emph{DPUs}) near the memory banks in the same DDR chip. 
HBM-PIM features 16-lane 16-bit \floatp SIMD units (called \emph{PCUs}) near the banks in the memory layers of an HBM stack~\cite{jedec.hbm.spec}. The SIMD units execute a only reduced set of instructions (e.g., multiplication, addition), since the architecture targets machine learning inference. 
AxDIMM with processing elements near the memory ranks is a DIMM-based solution. Its prototype places an FPGA fabric inside the buffer chip on the DIMM. The FPGA implements an accelerator for recommendation inference. 
AiM~\cite{lee2022isscc} is a GDDR6-based PIM architecture with near-bank processing units (\emph{PUs}) for multiply-and-accumulate and activation functions for deep learning applications. 
HB-PNM~\cite{niu2022isscc} is a 3D-stacked based PIM solution for recommendation systems. With hybrid bonding (HB) technology~\cite{fujun2020stacked}, HB-PNM stacks one layer of LPDDR4 DRAM~\cite{jedec2021lpddr4} on one logic layer. The logic layer embeds two types of specialized engines for \emph{matching} and \emph{ranking}, which the memory-bound steps of the evaluated recommendation system.

\begin{figure*}[h]
\centering
\includegraphics[width=1.0\linewidth]{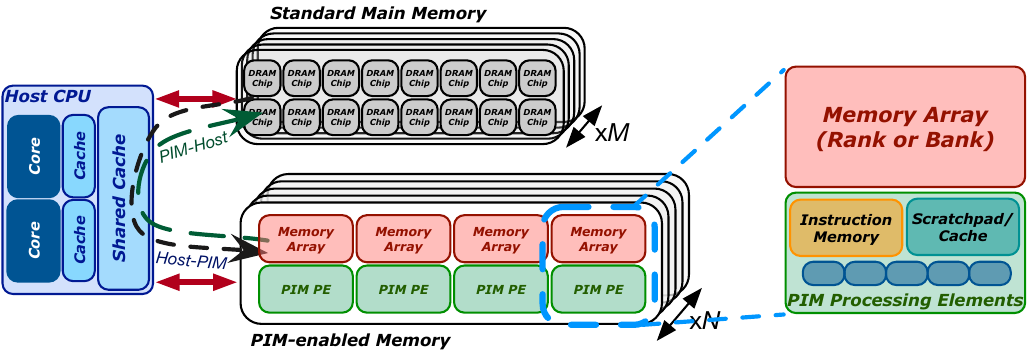}
\caption{High-level view of a state-of-the-art processing-in-memory system. {The host CPU has access to $M$ standard memory modules and $N$ PIM-enabled memory modules.}}
\label{fig:scheme}
\end{figure*}

{These five real-world PIM systems} 
have some important {common} characteristics, {as depicted in Figure~\ref{fig:scheme}}. 
First, there is a host processor (CPU or GPU), typically with a deep cache hierarchy, which has access to (1) standard main memory, and (2) PIM-enabled memory (i.e., UPMEM DIMMs, HBM-PIM stacks, AxDIMM DIMMs, AiM GDDR6, HB-PNM LPDDR4). 
Second, the PIM-enabled memory {chip} contains multiple PIM processing elements (PIM PEs), which have access to memory (either memory banks or ranks) with higher 
bandwidth {and lower latency} than the host processor. 
Third, the {PIM} processing elements (either {general-purpose} cores, SIMD {units, FPGAs, or specialized processors}) run at {only} a few hundred megahertz, and have a small number of registers and 
{relatively small} (or no) cache or scratchpad memory. 
Fourth, PIM PEs may not be able to communicate directly {with each other} (e.g., UPMEM DPUs, HBM-PIM PCUs or AiM PUs in different chips), 
{and} communication {between them} happens via the host processor. 
Figure~\ref{fig:scheme} shows a high-level view of such a state-of-the-art processing-in-memory system. 

In our study, we use the UPMEM PIM architecture~\cite{devaux2019, upmem, upmem2018, upmem-sdk, gomezluna2021benchmarking, upmem-guide}, the first PIM architecture to be commercialized in real hardware. 
The UPMEM PIM architecture uses conventional 2D DRAM arrays and combines them with general-purpose processing cores, called \emph{DRAM Processing Units} (\emph{DPUs}), on the same chip. 
In the current architecture generation (as of \juan{April 2023}), there are 8 DPUs and 8 DRAM banks per chip, and 16 chips per DIMM (8 chips/rank). 
DPUs are relatively deeply pipelined and fine-grained multithreaded~\cite{burtonsmith1978,smith1982architecture,thornton1970}. 
DPUs run software threads, called \emph{tasklets}, which are programmed in \emph{Single Program Multiple Data} (\emph{SPMD}) manner. 

DPUs have a 32-bit RISC-style general-purpose instruction set~\cite{upmem-guide}. They feature native support for \juan{32-bit integer addition/subtraction and 8-bit multiplication}, but some complex operations (e.g., 32-bit integer multiplication/division) and {\floatp} operations are emulated~\cite{gomezluna2021benchmarking, gomezluna2022ieeeaccess}. 

Each DPU has exclusive access to its own (1) 64-MB DRAM bank, called \emph{Main RAM} (\emph{MRAM}), (2) 24-KB instruction memory, and (3) 64-KB scratchpad memory, called \emph{Working RAM} (\emph{WRAM}). 
The host CPU can access the MRAM banks for copying input data (from main memory to MRAM) and retrieving results (from MRAM to main memory). These CPU-DPU/DPU-CPU transfers can be performed in parallel (i.e., concurrently across multiple MRAM banks), if the size of the buffers transferred from/to all MRAM banks is the same. Otherwise, the data transfers should be performed serially. Since there is no direct communication channel between DPUs, all inter-DPU communication takes place through the host CPU by using DPU-CPU and CPU-DPU data transfers.

Throughout this paper, we use generic terminology, since our implementation strategies are applicable to PIM systems like the generic one described in Figure~\ref{fig:scheme}, and not exclusive of the UPMEM PIM architecture. 
Thus, we use the terms \emph{PIM core}, \emph{PIM thread}, \emph{DRAM bank}, \emph{scratchpad}, and \emph{CPU-PIM/PIM-CPU transfer}, which correspond to DPU, tasklet, MRAM bank, WRAM, and CPU-DPU/DPU-CPU transfer in UPMEM's terminology~\cite{upmem-guide}.

\section{ML Training and PIM Implementation}
\label{sec:implementation}


We select four widely-used machine learning workloads (i.e., linear regression, logistic regression, \km clustering, and decision tree) as representative ones for our benchmarking and analysis of machine learning training on real-world processing-in-memory architectures. 
We consider them representative because they are diverse in terms of learning approach and application. They have also diverse computational characteristics (i.e., memory access pattern, computation pattern, synchronization needs), as Table~\ref{tab:diversity} shows.  

\begin{table*}[h]
\begin{center}
\caption{Machine learning workloads.}
\label{tab:diversity}
\resizebox{1.0\linewidth}{!}{
\begin{tabular}{|l|l|l|c||c|c|c|c|c|c|c|}
    \hline
    \textbf{Learning} & \multirow{2}{*}{\textbf{Application}} & \multirow{2}{*}{\textbf{Algorithm}} & \multirow{2}{*}{\textbf{Short name}} & \multicolumn{3}{c|}{\textbf{Memory access pattern}} & \multicolumn{2}{c|}{\textbf{Computation pattern}} & \multicolumn{2}{c|}{\textbf{Communication/synchronization}}  \\
    \cline{5-11}
     \textbf{approach} & & & & \textbf{Sequential} & \textbf{Strided} & \textbf{Random} & \textbf{Operations} & \textbf{Datatype} & \textbf{Intra PIM Core} & \textbf{Inter PIM Core}  \\
    \hline
    \hline
    \multirow{3}{*}{Supervised}
      & Regression & \textbf{Linear Regression} & \texttt{LIN} & Yes & No & No & mul, add & float, int32\_t & barrier & Yes \\
    \cline{2-11}
      & \multirow{2}{*}{Classification} & \textbf{Logistic Regression} & \texttt{LOG} & Yes & No & No & mul, add, exp, div & float, int32\_t & barrier & Yes \\
    \cline{3-11}
      & & \textbf{Decision Tree} & \texttt{DTR} & Yes & No & No & compare, add & float & barrier, mutex & Yes \\
    \hline
    \multirow{1}{*}{Unsupervised}
      & Clustering & \textbf{K-Means} & \texttt{KME} & Yes & No & No & mul, compare, add  & int16\_t, int64\_t & barrier, mutex & Yes \\
    \hline
\end{tabular}

}
\end{center}
\end{table*}


\subsection{Linear Regression}
\label{sec:linreg}
Linear regression~\cite{freedman2009statistical, yan2009linear} is a supervised learning algorithm where the predicted output variable has a linear relation with the input variable. 
Linear regression is widely used to model 
relationships between variables in biological and social sciences (e.g., epidemiology, environmental sciences, finance, etc.)~\cite{yan2009linear}. 

\noindent {\textbf{Algorithm Description.}} Linear regression obtains a linear model that predicts an output vector $y$ from an input matrix $X$ based on some coefficients or \emph{weights}, vector $w$. 
We implement linear regression with \emph{gradient descent}~\cite{polyak1987}, as the optimization algorithm to find the minimum of the loss function. 
During training, we repeatedly refine the values of vector $w$ based on the vector of observed values $y$ for the inputs in matrix $X$ (row vectors $x_i$). 
In each iteration, we first calculate the predicted output for each row vector $x_i$, i.e., the dot product of $x_i$ and $w$. 
Second, we calculate the gradient for the predicted output, i.e., we evaluate the error of the predicted output with respect to the observed value $y$. 
Third, we update the weights $w$ using the calculated gradient. 
We repeat the above process 
convergence (i.e., the gradient of loss function is zero or close to zero).

\noindent {\textbf{PIM Implementation.}} Our PIM implementation of linear regression with gradient descent divides the training dataset ($X$) so that each PIM core is assigned an equal number of row vectors $x_i$. 
If the training dataset resides initially in the main memory of the host processor, we need to transfer the corresponding partitions of the training dataset to the local memories (e.g., DRAM banks) of the PIM cores. 
Inside a PIM core, we proceed as follows. 
First, we further distribute the assigned row vectors $x_i$ across the running threads, which compute the dot products of row vectors and weights ($x_i \cdot w$). 
Second, each dot product result is compared to the observed value $y$ to compute a partial gradient value. 
Third, we reduce partial gradient values, and return the results to the host. 
Finally, the host (1) performs final reductions of gradient values from PIM cores, (2) updates the weights $w$, and (3) redistributes them to the PIM cores for the next training iteration. 

We implement four different versions of linear regression with different input datatypes and optimizations: (1) 32-bit \floatp (\texttt{LIN-FP32}), 
(2) 32-bit \fixedp (\texttt{LIN-INT32}), 
(3) \fixedp with hybrid precision (\texttt{LIN-HYB}), and (4) \fixedp with hybrid precision and built-in functions (\texttt{LIN-BUI}). 

\begin{itemize}
\item \texttt{LIN-FP32} trains with input datasets of real values (32-bit precision). 
\item \texttt{LIN-INT32} uses 32-bit \fixedp representation of input datasets. It uses 32-bit integer arithmetic. 
\item \texttt{LIN-HYB} is applicable to input datasets of a limited value range that can be represented in 8 bits. 
The dot product result is 16-bit width, and the final gradient is represented in 32 bits. 
This hybrid implementation is motivated by the fact that real-world PIM cores only feature arithmetic units of limited precision. For example, DPUs in the UPMEM PIM architecture~\cite{upmem-guide} run native 8-bit integer multiplication, but emulate 32-bit integer multiplication using \emph{shift-and-add} instructions~\cite{gomezluna2021benchmarking}. PCUs in HBM-PIM~\cite{kwon202125} and PUs in AiM~\cite{lee2022isscc} have 16-bit {\floatp} arithmetic units. 
\item \texttt{LIN-BUI} replaces compiler-generated 16-bit and 32-bit multiplications with a custom multiplication based on 8-bit built-in multiplication functions~\cite{upmem-sdk} (this optimization is specific to the UPMEM PIM architecture). {This optimization, which is based on the assumption that input data is encoded in 8 bits, reduces the number of instructions for each multiplication from 7 instructions (compiler-generated multiplication) to 4 (custom multiplication).}
Listing~\ref{lst:custom-mul} shows the default integer multiplication code (C-based (a) and compiled code (b)) and our custom integer multiplication code (C-based (c) and compiled code (d)).
\end{itemize}
%
%

\begin{figure}[h]
    \setcaptiontype{lstlisting}

    \begin{minipage}{\linewidth}
	    \begin{lstlisting}[style=myC]
%\HilightGray%result = X[i] * W[i]; // X and W are in WRAM (scratchpad)
        \end{lstlisting}
        \vspace{-8mm}
        \subcaption{Default integer multiplication: C-based code.}
        \label{sublst:codea}
	    \begin{lstlisting}[style=myC]
%\HilightYellow%lbs r3, r2, 0          // Load 1 byte from X[i]
%\HilightPink%lsl_add r2, r20, r1, 1 // Address of W[i]: r2=r20+(r1<<1)
%\HilightYellow%lhs r4, r2, 0          // Load 2 bytes from W[i]
%\HilightBlue%mul_ul_ul r2, r4, r3, small, 0x80000378 // r2=r4(l)*r3(l)
%\HilightBlue%mul_sh_ul r5, r4, r3      // r5=r4(h)*r3(l)
%\HilightGreen%lsl_add r2, r2, r5, 8     // r2=r2+(r5<<8)
%\HilightBlue%mul_sh_ul r5, r3, r4       // r5=r3(h)*r4(l)
%\HilightGreen%lsl_add r2, r2, r5, 8     // r2=r2+(r5<<8)
%\HilightBlue%mul_sh_sh r3, r4, r3       // r3=r4(h)*r3(h)
%\HilightGreen%lsl_add r2, r2, r3, 16, true, 0x80000378 //r2=r2+(r3<<16)
        \end{lstlisting}
        \vspace{-8mm}
        \subcaption{Default integer multiplication: Compiled code in UPMEM ISA.}
        \label{sublst:codea}
    \end{minipage}
   \vspace{2mm}
    \begin{minipage}{\linewidth}
	    \begin{lstlisting}[style=myC]
%\HilightBlue%__builtin_mul_sl_ul_rrr(templ, X[i], W[i]); 
%\HilightBlue%__builtin_mul_sl_sh_rrr(temph, X[i], W[i]); 
%\HilightGreen%result = (temph << 8) + templ; 
        \end{lstlisting}
        \vspace{-8mm}
        \subcaption{Custom integer multiplication: C-based code (with built-in functions).}
	    
	    \begin{lstlisting}[style=myC]
%\HilightYellow%lbs r4, r4, 0        // Load 1 byte from X[i]
%\HilightPink%lsl_add r5, r20, r3, 1 // Address of W[i]: r5=r20+(r1<<1)
%\HilightYellow%lhs r5, r5, 0        // Load 2 bytes from W[i]
%\HilightBlue%mul_sl_ul r6, r4, r5      // r6=r4(l)*r5(l)
%\HilightBlue%mul_sl_sh r4, r4, r5      // r4=r4(l)*r5(h)
%\HilightGreen%add r2, r6, r2           // r2=r2+r6
%\HilightGreen%lsl_add r2, r2, r4, 8    // r2=r2+(r4 << 8)
       \end{lstlisting}
       \vspace{-8mm}
       \subcaption{Custom integer multiplication: Compiled code in UPMEM ISA.}
       \label{sublst:codeb}
   \end{minipage}
   
\vspace{-2mm}    
    \caption{Default integer multiplication (C-based code (a) and compiled code (b)) vs. custom integer multiplication (C-based code (c) and compiled code (d)). The default multiplication compiles to 7 instructions (blue and green lines) while our custom multiplication compiles to 4 instructions.}
    \label{lst:custom-mul}
\end{figure}

In Section~\ref{sec:evaluation}, we evaluate all \texttt{LIN} versions in terms of accuracy (Section~\ref{sec:metrics}), performance for different numbers of threads per PIM core (Section~\ref{sec:analysis}), and performance scaling characteristics (Section~\ref{sec:scaling}). 
\juan{We also compare our \texttt{LIN} versions to 
\juang{custom} CPU and GPU implementations of linear regression (Section~\ref{sec:cpugpu}), which use Intel MKL~\cite{mkl} and NVIDIA cuBLAS~\cite{cublas}, respectively.}

\subsection{Logistic Regression}
\label{sec:logreg}
Logistic regression~\cite{freedman2009statistical, hosmer2013applied} is a supervised learning algorithm used for classification, which outputs probability values for each input observation variable or vector. This probability values represent the likelihood of belonging to a certain class or event. Logistic regression is used in various fields (e.g., medical, marketing, engineering, economics, etc.)~\cite{hosmer2013applied}. 

\noindent {\textbf{Algorithm Description.}} Logistic regression uses the \emph{sigmoid} function to map predicted values (output vector $y$ obtained from an input matrix $X$ and a weights vector $w$) to probabilities. 
Our implementation of logistic regression uses gradient descent, same as our linear regression implementation (Section~\ref{sec:linreg}). 
First, in the beginning of each training iteration, we obtain the dot product of row vectors $x_i$ and weights $w$. Second, we apply the sigmoid function to the dot product results. Third, we calculate the gradient to evaluate the error of the predicted probability. Fourth, we update the weights $w$ according to the gradients.

\noindent {\textbf{PIM Implementation.}} Our PIM implementation of logistic regression follows the same workload distribution pattern as our linear regression implementation. 
First, row vectors $x_i$ are distributed across PIM cores and threads in each PIM core. 
Second, each thread computes the dot product of a row vector and the weights ($x_i \cdot w$), and applies the sigmoid function to the dot product result. 
Third, the thread computes partial gradient values. 
Fourth, partial gradient values from different threads are reduced, and the results \juan{are returned} to the host. 
Finally, the host computes the final reductions, and updates the weights before redistributing them to the PIM cores. 

We implement six different versions of logistic regression with different input datatypes and optimizations: (1) 32-bit \floatp (\texttt{LOG-FP32}), (2) 32-bit \fixedp (\texttt{LOG-INT32}), (3) 32-bit \fixedp with LUT-based sigmoid calculation and LUT in DRAM (\texttt{LOG-INT32-LUT (MRAM)}), (4) 32-bit \fixedp with LUT-based sigmoid calculation and LUT in scratchpad (\texttt{LOG-INT32-LUT (WRAM)}), (5) \fixedp with hybrid precision and LUT-based sigmoid calculation (\texttt{LOG-HYB-LUT}), and (6) \fixedp with hybrid precision, LUT-based sigmoid calculation, and built-in functions (\texttt{LOG-BUI-LUT}).

\begin{figure}[h]
\centering
\includegraphics[width=1.0\linewidth]{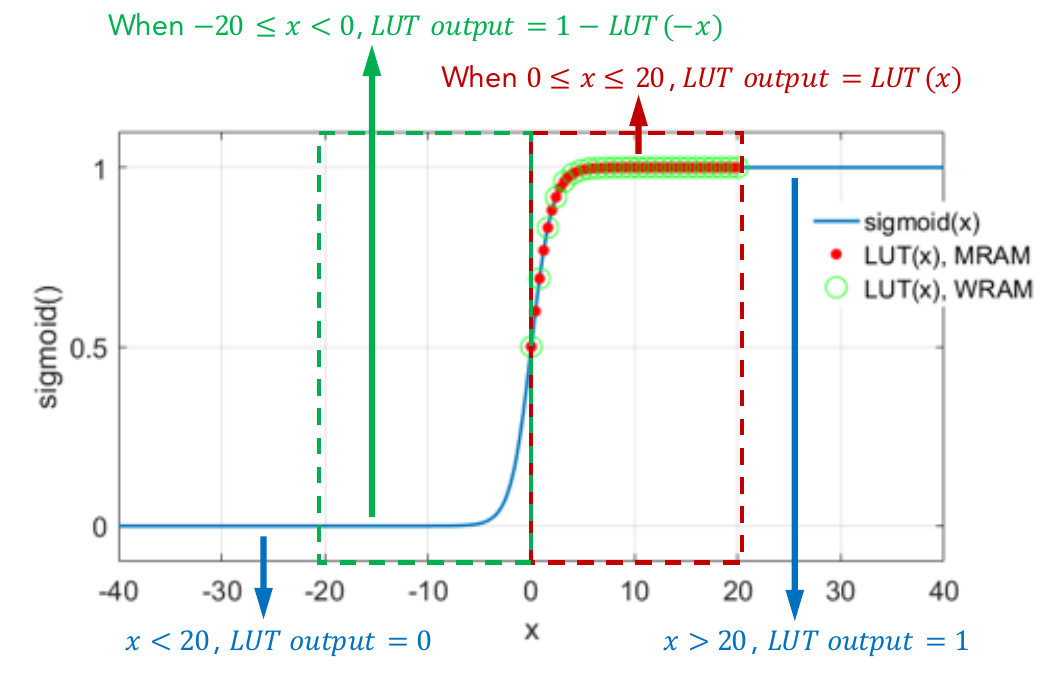}
\caption{LUT-based sigmoid calculation. The red dots represent the sigmoid values stored in the LUT.}
\label{fig:lut}
\end{figure}

\begin{itemize}
\item \texttt{LOG-FP32} trains with input datasets of real data (32-bit precision). If the PIM architecture does \emph{not} support exponentiation (needed for sigmoid), this operation can be approximated by Taylor series~\cite{weisstein2004taylor}. This is \juan{true for} the UPMEM PIM architecture. 
\item \texttt{LOG-INT32} uses 32-bit \fixedp representation of input datasets. It uses 32-bit integer arithmetic, and Taylor series for the sigmoid function. 
\item \texttt{LOG-INT32-LUT} versions use a lookup table (LUT) per PIM core for sigmoid values, instead of Taylor series. Figure~\ref{fig:lut} represents our LUT-based sigmoid calculation. The size of the LUT depends on the sigmoid boundary and the number of bits for the decimal part of the \fixedp representation. We take advantage of the fact that the sigmoid function is symmetric. Thus, for a sigmoid boundary of 20 and 10 bits for the decimal part, the size of the LUT is $20 \times 1024$ entries. To represent this range of values, we can fit the entries in 16 bits. As a result, the size of our LUT is 40 KB. 
This small size can comfortably reside in \juan{the} small scratchpads/caches of PIM cores (e.g., 64-KB WRAM in the UPMEM PIM architecture). However, occupying too much of a scratchpad may reduce the number of possible active threads~\cite{gomezluna2021benchmarking}. 
We analyze this version in Section~\ref{sec:analysis}, and compare it to a version that accesses the LUT directly from DRAM (e.g., MRAM in the UPMEM PIM architecture). Depending on where the LUT resides, we have two different versions: \texttt{LOG-INT32-LUT (MRAM)} and \texttt{LOG-INT32-LUT (WRAM)}. 
\item \texttt{LOG-HYB-LUT} is applicable to input datasets of a limited value range that can be represented in 8 bits, same as explained for \texttt{LIN-HYB}, and uses LUT-based sigmoid (LUT in scratchpad). 
\item \texttt{LOG-BUI-LUT} uses 8-bit builtin multiplication functions (Listing~\ref{lst:custom-mul}), same as explained for \texttt{LIN-BUI}, and uses LUT-based sigmoid (LUT in scratchpad).
\end{itemize}

In Section~\ref{sec:evaluation}, we evaluate all \texttt{LOG} versions in terms of accuracy (Section~\ref{sec:metrics}), performance for different numbers of threads per PIM core (Section~\ref{sec:analysis}), and performance scaling characteristics (Section~\ref{sec:scaling}).
\juan{We also compare our \texttt{LOG} versions to \juang{custom} CPU and GPU implementations of logistic regression (Section~\ref{sec:cpugpu}), \juan{which use Intel MKL~\cite{mkl}} and NVIDIA cuBLAS~\cite{cublas}, respectively.}

\subsection{Decision Tree}
\label{sec:dtree}
Decision trees~\cite{suthaharan2016decision} are tree-based methods used for classification and regression. They are frequently referred to as \emph{CART} (Classification and Regression Trees). 
A decision tree partitions the feature space into {\emph{leaves}},
with a simple prediction model in each leaf, 
typically a comparison to a threshold (e.g., an average value in regression problems, a majority class in classification problems).

\noindent {\textbf{Algorithm Description.}} The training process of a decision tree builds a binary-search tree, which represents the partitioning of the feature space. 
Each tree node splits the current rectangular sub-space further based on a feature and a threshold. 
The prediction is later done by following the correct path in the tree, up to a leaf which contains the predicted value. 

There are different flavors of decision tree algorithms, but the two main steps are typically:
\begin{enumerate}
\item Split a tree leaf, thus creating two children connected to their parent node (i.e., the \emph{old} leaf).
A split is represented as a tuple $(l, f, thresh)$, 
where $l$ is the tree leaf index, $f$ is the feature index, and $thresh$ is the feature threshold. 
After a split, 
the left child 
contains the points $p$ of the training set for which $p[f] <= thresh$, and the right child 
contains the points for which $p[f] > thresh$. 
\item Evaluate the quality of a tree leaf split. 
The quality of a split is measured with a specific score, e.g., the \emph{Gini impurity}~\cite{suthaharan2016decision}, a probability measure of a randomly chosen element being incorrectly labeled if it was randomly labeled.
\end{enumerate}

In this work, we implement decision trees for classification problems. Each feature vector of the training set is associated with a value referred to as its \emph{class}, and the goal of the training process is to create a tree which correctly predicts the class of previously unseen feature vectors. At each step, we choose one candidate threshold at random for every feature, and the best threshold-feature pair is used to generate the split. The resulting tree is known as \emph{extremely randomized tree}~\cite{geurts2006extremely}. This type of tree displays a larger bias and smaller variance than regular trees, and is meant to be used as the building block for forests of randomized trees~\cite{breiman2001random}.



\noindent {\textbf{PIM Implementation.}} Our PIM implementation of a decision tree partitions the training set into subsets of equal size, which the host processor transfers to the PIM cores. 
The host processor maintains the tree representation and makes \juan{splitting} decisions, while the PIM cores compute partial Gini scores to evaluate the splits. 
The partial Gini scores computed by PIM cores are returned to the host and aggregated, in order to make \juan{splitting} decisions based on the total Gini score.

The host maintains an active frontier of nodes, i.e., the current leaves of the tree. 
In each training iteration, the host decides whether (1) to split a tree leaf, an operation called \emph{split commit}, or (2) to evaluate a split, an operation called \emph{split evaluate}, or (3) to query the minimum and maximum values of a feature in a tree leaf, an operation called \emph{min-max}. The minimum value (\emph{min}) and the maximum value (\emph{max}) are needed by the host to randomly select a candidate split threshold in the [\emph{min}, \emph{max}] interval. 
Then, the host sends commands (i.e., split commit, split evaluate, min-max) to the PIM cores. The host can send multiple commands at once (with the only restriction that there must be at most one command per tree leaf), thus 
exploiting task-level parallelism in the PIM cores.



\begin{figure}[h]
\centering
\includegraphics[width=1.0\linewidth]{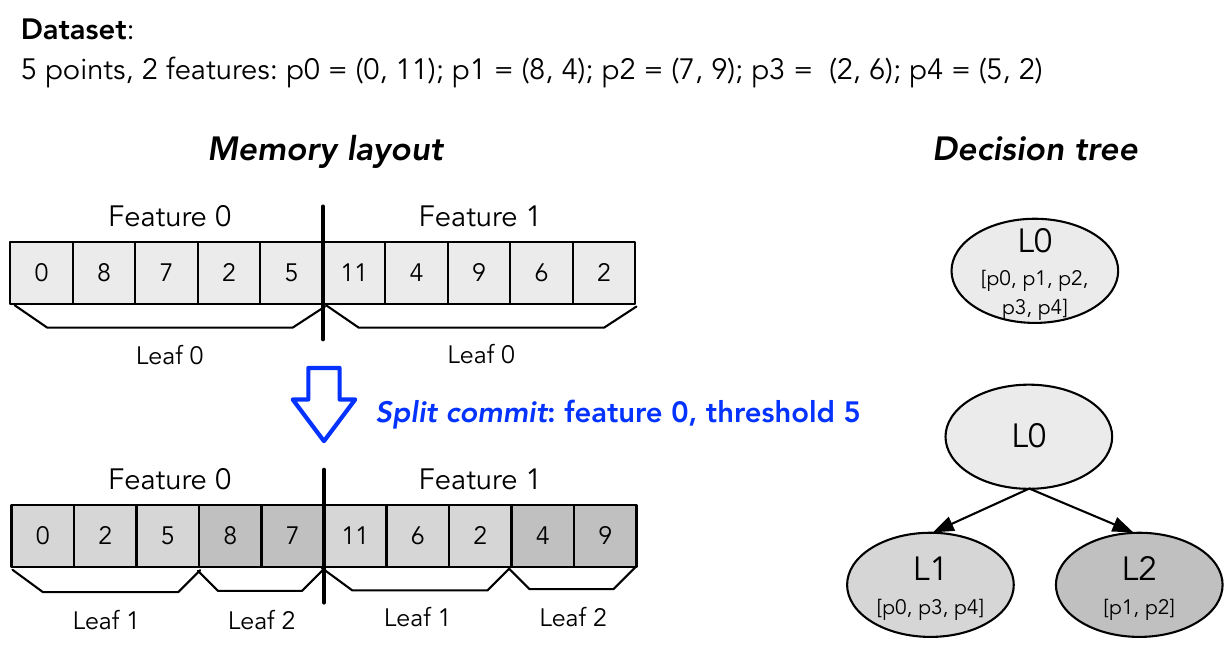}
\caption{Split commit operation in DTR.}
\label{fig:dtr}
\end{figure}

Inside a PIM core, a split evaluate command is also parallelized, as different PIM threads work on different batches of feature values. 
PIM threads move batches of feature values of the points in the training datasets from the DRAM bank to the scratchpad (i.e., from MRAM to WRAM in UPMEM DPUs), compare them to the corresponding threshold, and update the partial Gini score accordingly. 
This operation has low arithmetic intensity, since only one \floatp comparison and one integer addition are needed. 
Consequently, a key point for performance is to load and handle multiple feature values at once, in order to hide the latency of 
accesses to DRAM banks (e.g., in UPMEM DPUs, the MRAM-WRAM transfers are handled by a DMA engine with a 
deterministic cost for each transfer~\cite{gomezluna2021benchmarking}). 
Streaming memory accesses (using large MRAM-WRAM transfers in UPMEM DPUs) sustain higher memory bandwidth than fine-grained strided/random accesses (using short MRAM-WRAM transfers)~\cite{gomezluna2021benchmarking}. 
In order to access memory in streaming during split evaluate operations, we \juan{lay out} the training data in split commit operations as follows (see Figure~\ref{fig:dtr}): 
\begin{enumerate}
    \item Points are stored by features (leaf 0 in Figure~\ref{fig:dtr}). If we denote $p_i[f]$ the value of feature $f$ of point $p_i$, the first feature values are $p_0[0] p_1[0] ... p_n[0]$, then $p_0[1] p_1[1] ... p_n[1]$, etc.
    \item For all features, the feature values of points belonging to the same tree leaf are kept consecutive in memory (leaves 1 and 2 in Figure~\ref{fig:dtr}). This means that for a leaf node $l$ containing the subset of points $p^l_0, p^l_1, ..., p^l_k$, 
    and a feature $f$, the values of $p^l_0[f]p^l_1[f] ... p^l_k[f]$ are stored consecutively in memory. The same applies to the class values. 
\end{enumerate}

Figure~\ref{fig:dtr} illustrates the split commit operation on a dataset of 5 points with 2 features. Initially (L0), the points are stored in memory with $p_0-p_4$’s feature 0 values, and then feature 1 values. 
After the split commit of L0 on feature 0 with threshold 5, two new leaves are created, L1 with $p_0, p_3, p_4$ (feature 0 value <= 5) and L2 with $p_1, p_5$ (feature 0 value > 5). 
The split commit operation reorders the values\footnote{The reordering is partial in the sense that we only need to place points of the same leaf together, not to sort them.} in memory, so that points in the same leaf are stored consecutively. This way, PIM threads can perform streaming memory accesses during the split evaluate operation. 
The cost of reordering is largely compensated by the benefits of this data layout, since split commit is less frequent than split evaluate. 
The reordering is a parallel operation where a different PIM thread reorders a different feature.

In Section~\ref{sec:evaluation}, we evaluate \texttt{DTR} in terms of accuracy (Section~\ref{sec:metrics}), performance for different numbers of threads per PIM core (Section~\ref{sec:analysis}), and performance scaling characteristics (Section~\ref{sec:scaling}).
\juan{We also compare our \texttt{DTR} implementation to state-of-the-art CPU and GPU implementations of decision tree (Section~\ref{sec:cpugpu}). 
The CPU version is from Scikit-learn~\cite{pedregosa2011scikit} and the GPU version \juang{is} from RAPIDS~\cite{rapids}.}


\subsection{\km Clustering}
\label{sec:kmeans}

\km~\cite{Lloyd82leastsquares} is a classic iterative clustering method used to find groups which have not been explicitly labeled in a dataset. 
\km can be used to identify unknown groups in complex multidimensional datasets useful to business data segmentation,  and with a prominent use in image processing in an effort of simplifying images ahead of more complex classification models and compression algorithms. 

\noindent {\textbf{Algorithm Description.}} A \km algorithm attempts to partition the dataset into K pre-defined distinct non-overlapping subgroups (\emph{clusters}) where each data point belongs to only one group. Points within a cluster are meant be as similar (close) as possible while in comparison to points belonging to other clusters, their differences (\emph{distance}) should be maximized. A cluster is identified 
by its \emph{centroid}, a point with coordinates determined as the minimum total distance between itself and each point of the cluster. 
Our \km algorithm follows Lloyd's method~\cite{Lloyd82leastsquares}. 

\noindent {\textbf{PIM Implementation.}} Our PIM implementation 
of \km partitions the training set and distributes it evenly over the PIM cores. 
The host processor sets initial random values of the centroids and broadcasts them to all PIM cores. 
The algorithm 
follows an iterative process in which (1) each PIM core assigns points of its 
part of the training set to the clusters, and then (2) the 
host processor adjusts the centroids based on the new assignment of points.

First, inside a PIM core, PIM threads evaluate which centroid is the nearest one to each point of the training set. 
Distance calculations are done using 16-bit integer arithmetic. 
Input data are quantized over a range of $\pm32767$ (16-bit signed integers) to avoid overflowing when doing summations. 
Second, after finding the nearest centroid to a point, a PIM thread increments a counter and updates one accumulator per coordinate. The counter and the accumulators are associated to the corresponding cluster. Each per-coordinate accumulator contains the sum of values of the corresponding coordinate for all points belonging to a cluster. 
After all points are processed, each PIM core has partial sums of the coordinate values of the points in each cluster, and the number of points in each cluster. 
Third, the host processor then retrieves all per-cluster partial sums and counts from all PIM cores, and reduces them in order to compute the new coordinates of the centroids (calculated as the total sum of each coordinate divided by the total count). 
If these new centroid coordinates are far enough from the previous ones, they are sent over to the PIM cores for another iteration. 
The process continues until a centroid's coordinates \juan{converge} to a local optimum, i.e., when the updated coordinates are within a threshold distance to the previous coordinates. 
The threshold distance used to check \juan{for} convergence is the \emph{Frobenius norm}~\cite{weissteinFrobenius}. 
Fourth, once a clustering is completed, the PIM cores computes the \emph{inertia} (also known as \emph{within-cluster sum-of-squares}) of the clustering for their assigned points, and the host processors sums them up. The entire \km algorithm is repeated with different random starting centroids. The host processor \juan{chooses} the clustering with the lowest inertia as the final result.



In Section~\ref{sec:evaluation}, we evaluate \texttt{KME} in terms of quality and similarity (Section~\ref{sec:metrics}), performance for different numbers of threads per PIM core (Section~\ref{sec:analysis}), and performance scaling characteristics (Section~\ref{sec:scaling}).
\juan{We also compare our \texttt{KME} implementation to state-of-the-art CPU and GPU implementations of \km (Section~\ref{sec:cpugpu}). 
The CPU version is from Scikit-learn~\cite{pedregosa2011scikit} and the GPU version \juang{is} from RAPIDS~\cite{rapids}.}

\ignore{
\julien{Julien: I think the following enumeration is repeating what is said previously, maybe we drop it ?} \jgl{Agree.}

		   Initialisation phase (host):

\begin{enumerate}
\item The first step consists in equally splitting and filling up each of the DPU's memory banks with random points from the dataset. The K centroids are randomly initialized. 
\item The coordinates of the K centroids are broadcasted to all DPUs.

 DPU execution:

\item Initialisation of the cluster sum and the cluster counter.
\item For each point, measure the distance to each centroid and determine the closest one.
add features to the appropriate cluster sum and increment the cluster counter.

CPU centralization and loop:

\item Gather all sums and counters for every DPUs

\item Compute the new centroid positions

\item Compare to old position and start a new DPU run if above the threshold (loop back in 2.)

\item Compute the points clustering based on the final position of the centroids 

\end{enumerate}}


\ignore{
\jgl{The figure and the algorithms look good, but I think we do not need them in the current iteration of the paper.}

\begin{center}
\resizebox{4cm}{8cm}{
\begin{tikzpicture}[node distance=0.8cm]
\node (in1) [io] {Load points to DPU memory};
\node (pro1c) [io, below=\vspacing of in1] {Broadcast centroids to all DPUs};
\node (pro1) [processdpu, below=\vspacing of pro1c] {Initialize cluster sums and counters to 0};
\node (pro2) [processdpu, below=\vspacing of pro1] {Find the nearest centroid};
\node (pro3) [processdpu, below=\vspacing of pro2] {update cluster sum and counter};
\node (pro2c) [processcpu, below=\vspacing of pro3] {Agglomerate cluster sums and counters from all DPUs};
\node (pro3c) [processcpu, below=\vspacing of pro2c] {Average the new centroids of clusters};
\node (dec1) [decision, below=\vspacing of pro3c] {Centroids changed?};
\node (out1) [stop, below=\vspacing of dec1] {Output centroids};

\draw [arrow] (in1) -- (pro1c);
\draw [arrow] (pro1c) -- (pro1);
\draw [arrow] (pro1) -- (pro2);
\draw [arrow] (pro2) -- (pro3);
\draw [arrow] (pro3) -- (pro2c);
\draw [arrow] (pro2c) -- (pro3c);
\draw [arrow] (pro3c) -- (dec1);
\draw [arrow] (dec1) -- node[anchor=east] {no} (out1);
\draw [arrow] (dec1.east) -- node[anchor=south] {yes} ++(3,0) |- (pro1c.east);

\background{pro1}{pro1}{pro3}{pro3}{I}
\end{tikzpicture}
}
\end{center}
}

\ignore{
\makeatletter
\algnewcommand{\LineComment}[1]{\Statex \hskip\ALG@thistlm \(\triangleright\) #1}
\algnewcommand{\IndentLineComment}[1]{\Statex \hskip\ALG@tlm \(\triangleright\) #1}
\makeatother

\begin{algorithm}
\footnotesize
\caption{KMeans on DPU - host side}
\textbf{Parameters:}   \\
\hspace*{\algorithmicindent} $N$ \Comment{ number of data points }\\
\hspace*{\algorithmicindent} $K$ \Comment{ number of clusters }\\
\hspace*{\algorithmicindent} $tol$ \Comment{ convergence criterion} \\
\hspace*{\algorithmicindent} $D$ \Comment{ number of DPU} \\
\textbf{Input:} \\
\hspace*{\algorithmicindent}  $Centroids_{old}=\{ o_1, \ldots, o_K \}$ \Comment{Initial centroids} \\
\hspace*{\algorithmicindent}  $points=\{ p_1, \ldots, p_N\}$ \Comment{Data points} \\
\textbf{Output:} \\
\hspace*{\algorithmicindent} $Centroids=\{c_1, \ldots, c_K \}$ \Comment{Final centroids} \\
\hspace*{\algorithmicindent} $\mathcal{P} = \{ \mathcal{C}_1, \ldots, \mathcal{C}_K \}$ \Comment{A partition of $points$}
\begin{algorithmic}[1]
\Require $N \geq K$
    \State $N_D \gets N / D$ \Comment{number of points per DPU}
    \LineComment{Fill the DPU memory}
    \For{$i \gets 0,D-1$}
        \State $points_{DPU i} \gets \{p_{i \cdot N_D + 1}, \ldots, p_{(i+1) \cdot N_D} \}$
    \EndFor
    \While{$d_F (Centroids, Centroids_{old} ) > tol$} \Comment{$d_F$: Frobenius distance}
        \IndentLineComment{Broadcast current centroids to the DPUs}
        \For{$i \gets 0,N-1$}
            \State $Centroids_{DPU i} \gets  Centroids$
        \EndFor
\algstore{bkbreak}
\end{algorithmic}
\end{algorithm}

\begin{algorithm}
\addtocounter{algorithm}{-1}
\footnotesize
\caption{KMeans on DPU - DIMM side}
\begin{algorithmic}[1]
\algrestore{bkbreak}
        \State $Clusters\_sum = \{ s_1, \ldots, s_K \} \gets \{\mathbf{0}, \ldots, \mathbf{0}\}$
        \State $Clusters\_size = \{ sz_1, \ldots, sz_K \} \gets \{0, \ldots, 0\}$
        \For{$i \gets 1,N_D$}
            \IndentLineComment{\textbf{Assignment step:}}
            \State $d_{min} = \infty$
            \For{$j \gets 1,K$}
                \State $d \gets \lvert p_i - c_j \rvert$
                \If{$d < d_{min}$}
                    \State $d_{min} \gets d$
                    \State $n \gets j$
                \EndIf
            \EndFor
            \LineComment{\textbf{Update step:}}
            \State $s_n \gets s_n + p_i$
            \State $sz_n \gets sz_n + 1$
        \EndFor
\algstore{bkbreak}
\end{algorithmic}
\end{algorithm}

\begin{algorithm}
\addtocounter{algorithm}{-1} 
\footnotesize
\caption{KMeans on DPU - host side}
\begin{algorithmic}[1]
\algrestore{bkbreak}
    \State $Clusters\_sum \gets \sum\nolimits_{DPU} Clusters\_sum_{DPU}$
    \State $Clusters\_size \gets \sum\nolimits_{DPU} Clusters\_size_{DPU}$
    \State $Centroids_{old} \gets Centroids$
    \For{$i \gets 1,K$}
        \State $c_i \gets s_i / sz_i$
    \EndFor
    \EndWhile
    \State $\mathcal{C}_i \gets \{ p_j \mid \forall h \in \llbracket 1..N \rrbracket : d(p_j, c_i) \leq d(p_j, c_h) \}$
    \State\Return $Centroids, \mathcal{P}$
\end{algorithmic}
\end{algorithm}
 
\begin{algorithm}
\footnotesize
\caption{KMeans on CPU}\label{alg:blabla}
\textbf{Parameters:}   \\
\hspace*{\algorithmicindent} $N$ \Comment{ number of data points }\\
\hspace*{\algorithmicindent} $K$ \Comment{ number of clusters }\\
\hspace*{\algorithmicindent} $tol$ \Comment{ convergence criterion} \\
\textbf{Input:} \\
\hspace*{\algorithmicindent}  $Centroids_{old}=\{ o_1, \ldots, o_K \}$ \Comment{Initial centroids} \\
\hspace*{\algorithmicindent}  $points=\{ p_1, \ldots, p_N\}$ \Comment{Data points} \\
\textbf{Output:} \\
\hspace*{\algorithmicindent} $Centroids=\{c_1, \ldots, c_K \}$ \Comment{Final centroids} \\
\hspace*{\algorithmicindent} $\mathcal{P} = \{ \mathcal{C}_1, \ldots, \mathcal{C}_K \}$ \Comment{A partition of $points$}
\begin{algorithmic}[1]
\Require $N \geq K$
    \While{$d_F (Centroids, Centroids_{old} ) > tol$} \Comment{$d_F$: Frobenius distance}
        \IndentLineComment{\textbf{Assignment step:} assign each point to the cluster with the nearest centroid}
        \State $\mathcal{C}_i \gets \{ p_j \mid \forall h \in \llbracket 1..N \rrbracket : d(p_j, c_i) \leq d(p_j, c_h) \}$
        \LineComment{\textbf{Update step:} update the centroids}
        \State $Centroids_{old} \gets Centroids$
        \For {$i \gets 1,K$}
            \State $\displaystyle c_i = \frac{1}{\lvert \mathcal{C}_i \rvert} \sum\nolimits_{p_j \in \mathcal{C}_i} p_j$
        \EndFor
    \EndWhile
    \State $\mathcal{C}_i \gets \{ p_j \mid \forall h \in \llbracket 1..N \rrbracket : d(p_j, c_i) \leq d(p_j, c_h) \}$
    \State\Return $Centroids, \mathcal{P}$
\end{algorithmic}
\end{algorithm}
}


\section{Methodology}
\label{sec:methodology}
We make our implementations of ML workloads for a real-world PIM system compatible with  Scikit-learn~\cite{pedregosa2011scikit}, an open-source machine learning library, by deploying them as Scikit-learn estimator objects. 

We run our experiments on a real-world PIM system~\cite{upmem} with 2,524 PIM cores running at 425 MHz, and 158 GB of DRAM memory.\footnote{The UPMEM-based PIM system can have up to 2,560 PIM cores and 160 GB of DRAM.} 
Table~\ref{tab:pim-cpugpu} shows the main characteristics of this PIM system. The table also includes characteristics of the CPU and the GPU that we use as baselines for comparison (Section~\ref{sec:cpugpu}).
\juan{We compare our PIM implementations of ML workloads to state-of-the-art CPU and GPU implementations of the same workloads in terms of performance and quality (Section~\ref{sec:cpugpu}). 
For linear and logistic regression, we implement CPU versions with Intel MKL~\cite{mkl} and GPU versions with NVIDIA cuBLAS~\cite{cublas}. For decision tree and K-Means, CPU versions are from Scikit-learn~\cite{pedregosa2011scikit} and GPU versions from RAPIDS~\cite{rapids}.}

\begin{table*}[h]
\begin{center}
\caption{Evaluated PIM system, baseline CPU, and baseline GPU.}
\label{tab:pim-cpugpu}
\resizebox{1.0\linewidth}{!}{
\begin{tabular}{|l||c|ccc|cc|c|}
\hline
\multirow{2}{*}{\textbf{System}} & \textbf{Process} & \multicolumn{3}{c|}{\textbf{Processor Cores}} & \multicolumn{2}{c|}{\textbf{\juang{Main} Memory}} & \multirow{2}{*}{\textbf{TDP}} \\
\cline{3-7}
& \textbf{Node} & \textbf{Total Cores} & \textbf{Frequency} & \textbf{Peak Performance} & \textbf{Capacity} & \textbf{Total Bandwidth} & \\
\hline
\hline
\textbf{UPMEM PIM System}~\cite{upmem} & 2x nm & \juan{2,524}$^\diamond$ & 425 MHz & 1,088 GOPS & \juan{158} GB & \juan{2145 GB/s} & 280 W$^\dagger$ \\ \hline
\textbf{Intel Xeon Silver 4215 CPU}~\cite{xeon-4215} & 14 nm & 8 (16 threads) & 2.5 GHz & 40 GFLOPS$^\star$ & 256 GB & 37.5 GB/s & 85 W \\ \hline
\textbf{NVIDIA A100 GPU}~\cite{a100} & 7 nm & 108 (6,912 SIMD lanes) & 1.4 GHz & 19,500 GFLOPS & 40 GB & 1555 GB/s & 250 W \\ \hline
\end{tabular}

}
\end{center}
\begin{flushleft}
\resizebox{0.5\linewidth}{!}{
$\begin{tabular}{l}
$^\diamond$ There are several faulty PIM cores in the PIM system where we run our experiments. \\
$^\dagger Estimated\ TDP = \frac{Total\ PIM\ cores}{PIM\ cores/DIMM} \times 14\ W/DIMM$~\cite{upmem}. \\
$^\star Estimated\ GFLOPS = 2.5\ GHz \times 8\ cores \times 2\ instructions\ per\ cycle$.
\end{tabular}$
}
\end{flushleft}
\end{table*}

Table~\ref{tab:datasets} presents the datasets that we use in different experiments. 
For analysis of PIM kernel performance and performance scaling (both weak and strong scaling) experiments (Sections~\ref{sec:analysis} and~\ref{sec:scaling}), we use synthetic datasets, since we can generate them as large as needed for the scaling experiments. 
For comparison to CPU and GPU (Section~\ref{sec:cpugpu}), we use state-of-the-art real datasets. 
For \texttt{LIN}, we use the SUSY dataset~\cite{baldi2014searching} available at~\cite{susy}. This dataset contains 5 million samples with 18 \floatp attributes. 
For \texttt{LOG}, we use the SUSY dataset and the Skin segmentation dataset~\cite{skin}. This dataset contains 245,057 samples with 3 
integer attributes. 
For \texttt{DTR} and \texttt{KME}, we use the Higgs boson dataset~\cite{baldi2014searching} available at~\cite{Dua:2019}. This dataset consists of 11 million points with 28 \floatp features, and one binary target label. For DTR, the last 500,000 points are used as the test set. For KME, the whole set is used for clustering. 
\juangg{For \texttt{DTR} and \texttt{KME}, we also use the Criteo 1TB Click Logs dataset~\cite{criteo}. The entire dataset contains 24 days of feature values and click feedback for millions of display ads. 
Samples in this dataset have 40 attributes. 
Given the huge size of the entire dataset, we only use parts of it in two different experiments. 
In the first experiment, we use one quarter of day 0 (49 million samples). This is the maximum size that we can use in the GPU that we use in our experiments (Table~\ref{tab:pim-cpugpu}). 
In the second experiment, we use 2 days (days 0 and 1, i.e., 395 million samples). We can only run this experiment on the CPU and the PIM system.}

\begin{table*}[h]
\begin{center}
\caption{Datasets.}
\label{tab:datasets}
\centering
\resizebox{\linewidth}{!}{
\begin{tabular}{|l||l|l|l|}
\hline
\multirow{2}{*}{\textbf{ML Workload}} & \multicolumn{2}{c|}{\textbf{Synthetic Datasets}$^\dagger$} & \multirow{2}{*}{\textbf{Real Datasets}} \\
\cline{2-3}
  & \textbf{Strong Scaling (1 PIM core | 256-2048 PIM cores)} & \textbf{Weak Scaling (per PIM core)} &   \\
\hline
\hline
Linear regression & 2,048 samples, 16 attr. (0.125 MB) | 6,291,456 samples, 16 attr. (384 MB) & 2,048 samples, 16 attr. (0.125 MB) & SUSY~\cite{baldi2014searching, susy} \\ \hline
Logistic regression & 2,048 samples, 16 attr. (0.125 MB) | 6,291,456 samples, 16 attr. (384 MB) & 2,048 samples, 16 attr. (0.125 MB) & Skin segmentation~\cite{skin} \\ \hline
Decision tree & 60,000 samples, 16 attr. (3.84 MB) | 153,600,000 samples, 16 attr. (9830 MB) & 600,000 samples, 16 attr. (38.4 MB) & Higgs boson~\cite{baldi2014searching, Dua:2019} \juangg{| Criteo~\cite{criteo}}\\ \hline
K-Means & 10,000 samples, 16 attr. (0.64 MB) | 25,600,000 samples, 16 attr. (1640 MB) & 100,000 samples, 16 attr. (6.4 MB) & Higgs boson~\cite{baldi2014searching, Dua:2019} \juangg{| Criteo~\cite{criteo}}\\ \hline
\end{tabular}

}
\end{center}
\begin{flushleft}
\resizebox{0.38\linewidth}{!}{
$\begin{tabular}{l}
$^\dagger$ Format = Samples \juan{(dataset elements)}, Attributes (Size in MB).
\end{tabular}$
}
\end{flushleft}
\end{table*}




\subsection{\juan{ML Training Quality} Metrics}
\label{sec:metrics}

We evaluate the \juan{training} quality of the different versions of our ML workloads. 
We use synthetic datasets \juan{(with uniformly distributed random samples)} and run the experiments on a single PIM core (i.e., an UPMEM DPU). 


For \texttt{LIN} and \texttt{LOG}, the original synthetic datasets contain samples \juan{(i.e., dataset elements, each with a number of attributes)} with 4 decimal numbers, represented as 32-bit \floatp values. The fixed point versions use the same datasets after quantization. The number of samples is 8,192 and the number of attributes is 16. 
We calculate the \emph{training error rate} \juan{(lower is better)} as the percentage of inference errors of a model for the same data the model was trained on. 

For \texttt{DTR}, the synthetic dataset has 32-bit \floatp values. The data is \emph{not} quantized. The number of samples is 600,000 and the number of attributes is 16. There are 4 informative attributes, 4 redundant attributes (a random linear combination of the informative attributes), and 8 random attributes. We evaluate the \juan{\emph{training accuracy} (closer to 1 is better)} on the same data the model was trained on.

For \texttt{KME}, the synthetic dataset has 32-bit \floatp values. The PIM version uses the same dataset after quantization. The number of samples is 100,000 and the number of attributes is 16. Because it is an unsupervised problem, we do \emph{not} use accuracy as a metric. Instead, we use the \emph{Calinski-Harabasz score}~\cite{calinski1974dendrite} to measure the absolute quality of the clustering with no knowledge of the ground truth used in the generation of the dataset. We also measure the similarity of the clusterings produced by the Scikit-learn version of the algorithm with the PIM implementation using the \emph{adjusted Rand index}~\cite{hubert1985comparing}.

\section{Evaluation}
\label{sec:evaluation}

\juan{In this section, we evaluate our implementations of ML workloads (Section~\ref{sec:implementation}) on a real-world PIM system~\cite{upmem}, 
according to our evaluation methodology (Section~\ref{sec:methodology}). 
First, we evaluate the {quality} of our implementations (Section~\ref{sec:metrics}). 
Second, we analyze the performance of the different versions of our ML workloads on a single PIM core for different numbers of PIM threads (Section~\ref{sec:analysis}). 
Third, we evaluate the performance scaling characteristics of our ML workloads on the PIM system (Section~\ref{sec:scaling}). 
Fourth, we compare the performance and energy consumption of our implementations for the PIM system to their state-of-the-art CPU and GPU counterparts (Section~\ref{sec:cpugpu}).} 

\subsection{\juang{ML Training Quality}}
\label{sec:evaluation:accuracy}

\subsubsection{\juang{Linear Regression (LIN)}}
\begin{figure}[h]
\centering
\includegraphics[width=0.9\linewidth]{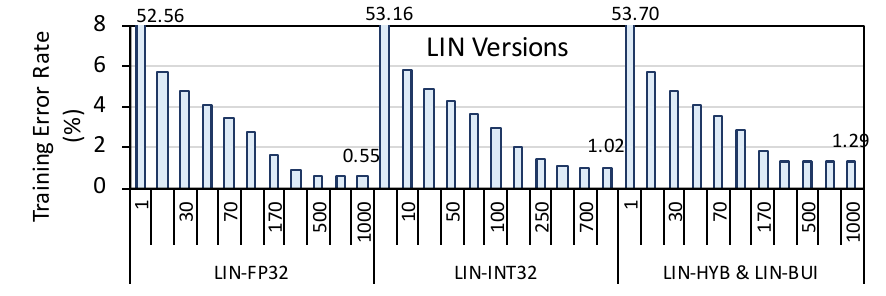}
\caption{Training error rate (\%) of \texttt{LIN} versions.}
\label{fig:LIN-acc}
\end{figure}

Figure~\ref{fig:LIN-acc} shows the training error rate of our four versions of \texttt{LIN} for \juang{varying} numbers of training iterations between 1 and 1000. 
We observe that the training error rate flattens after 500 iterations for the four versions.  
\texttt{LIN-FP32} achieves a training error rate as low as 0.55\% (same as the CPU version). This is the comparison point for the integer versions (i.e., \texttt{LIN-INT32}, \texttt{LIN-HYB}, \texttt{LIN-BUI}). 
The training error rate of the integer versions remains low {(1.02\% for \texttt{LIN-INT32} and 1.29\% for \texttt{LIN-HYB} and \texttt{LIN-BUI})} and close to that of the 32-bit {\floatp} version, as shown in the figure. 
\texttt{LIN-HYB} and \texttt{LIN-BUI} show the same behavior, since they use the same datatypes.

\subsubsection{\juang{Logistic Regression (LOG)}}
\begin{figure}[h]
\centering
\includegraphics[width=1.0\linewidth]{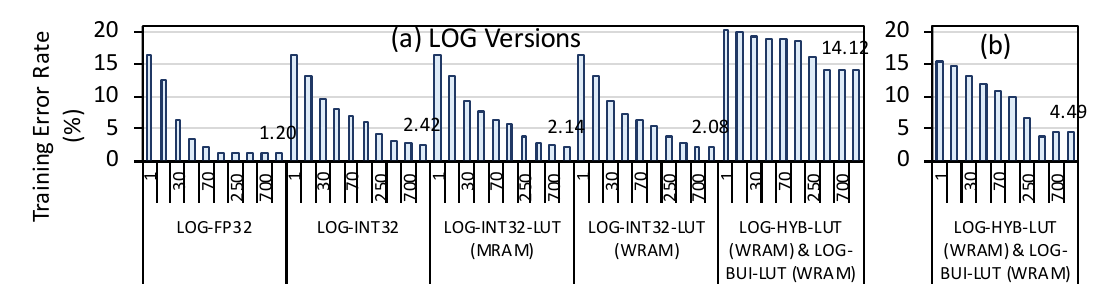}
\caption{Training error rate (\%) of \texttt{LOG} versions.}
\label{fig:LOG-acc}
\end{figure}

Figure~\ref{fig:LOG-acc}(a) presents the training error rate of our six versions of \texttt{LOG} for numbers of training iterations between 1 and 1000. 
The training error of \texttt{LOG-FP32}, which we use as the comparison point for the integer versions (i.e., \texttt{LOG-INT32}, \texttt{LOG-INT32-LUT (MRAM)}, \texttt{LOG-INT32-LUT (WRAM)}, \texttt{LOG-HYB-LUT (WRAM)}, \texttt{LOG-BUI-LUT (WRAM)}), is almost flat after 100 iterations, and is as low as 1.20\% after 1000 iterations (same as the CPU version). 
We observe that the training error rate of \texttt{LOG-INT32} {(2.42\%)} is higher than that of \texttt{LOG-INT32-LUT (MRAM)} and \texttt{LOG-INT32-LUT (WRAM)} {(2.14\%)}. 
The reason is that \texttt{LOG-INT32} approximates exponentiation (hence, sigmoid) with Taylor series, while \texttt{LOG-INT32-LUT (MRAM)} and \texttt{LOG-INT32-LUT (WRAM)} store exact sigmoid values in a LUT. 
\texttt{LOG-HYB-LUT (WRAM)} and \texttt{LOG-BUI-LUT (WRAM)} increase the training error rate significantly (14.12\%) due to the use of reduced-precision datatypes \juang{(i.e., 8- and 16-bit integers)}. 
In another experiment using samples with 2 decimal numbers (Figure~\ref{fig:LOG-acc}(b)), the training error rate of these two versions decreases to 4.49\%.

\subsubsection{\juang{Decision Tree (DTR)}}
We limit the tree depth to 10. The tree is built by splitting leaf nodes until no node can be split. A node cannot be split if it holds fewer than two data points, or if it contains only points belonging to the same class, or if its depth exceeds the maximum tree depth. To account for the effect of different \juang{synthetic datasets (with randomly generated samples) on both PIM and CPU} implementations, we restart the algorithm 10 times, and average the resulting accuracies. We register a training accuracy of 0.90008 for the PIM implementation, against 0.90175 for the CPU version.

\subsubsection{\juang{\km Clustering (KME)}}
We perform a \km clustering with 16 clusters to match the dataset generation. The clustering iterates for a maximum of 300 iterations, or until the relative Frobenius norm between the cluster centers of two consecutive iterations is lower than 0.0001. In practice, the clustering always converges after less than 40 iterations on both the PIM and CPU implementations. To account for \juang{the effect of synthetic datasets (with randomly generated samples)}, we average the 
metrics on 10 runs with different random seeds. We register an average Calinski-Harabasz scores~\cite{calinski1974dendrite} of 82200 for both implementations. The adjusted Rand index~\cite{hubert1985comparing} between the PIM and CPU clusterings is 0.999347 on average, showing that the clusterings are nearly identical despite the quantization.

\subsection{Performance Analysis of PIM Kernels}
\label{sec:analysis}
We analyze in this section the performance of the different PIM kernel versions of our ML workloads on a single PIM core (i.e., an UPMEM DPU). 
This way, we understand the effect of (1) different optimizations we apply, and (2) increasing the number of PIM threads. 

\subsubsection{\juang{Linear Regression (LIN)}}
\begin{figure}[h]
\centering
\includegraphics[width=1.0\linewidth]{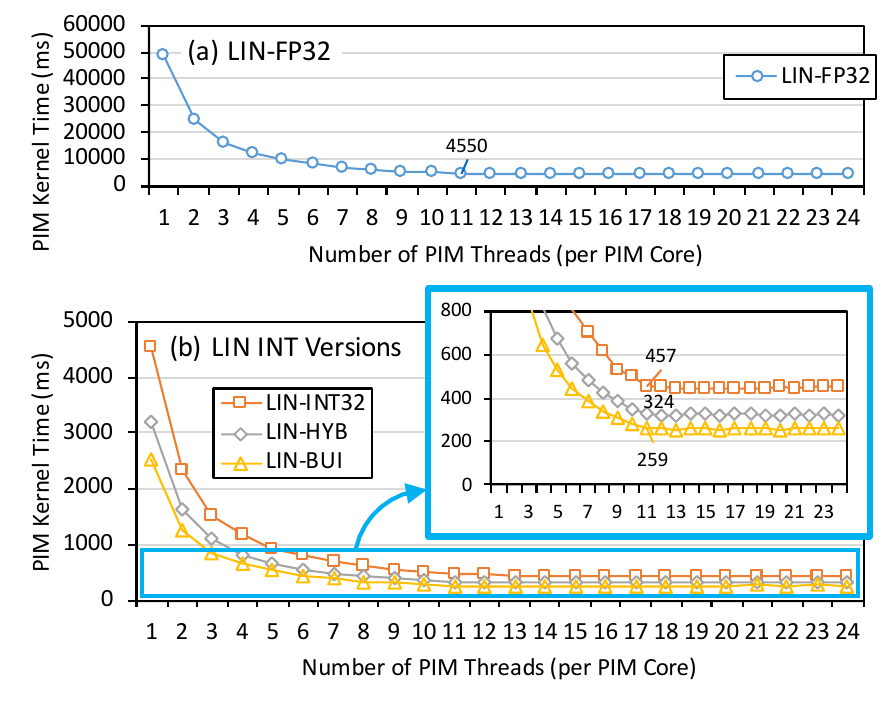}
\caption{Execution time (ms) of four versions of linear regression \juang{using} 1-24 PIM threads in 1 PIM core.}
\label{fig:lin-1dpu}
\end{figure}

Figure~\ref{fig:lin-1dpu} shows the PIM kernel time of our four versions of \texttt{LIN}. 
The upper plot (Figure~\ref{fig:lin-1dpu}(a)) represents the PIM kernel time of \texttt{LIN-FP32}. 
The lower plot (Figure~\ref{fig:lin-1dpu}(b))  shows the PIM kernel time of the integer versions. 
We make four observations. 
First, all \texttt{LIN} versions result in their best performance with 11 or more PIM threads. Eleven is the minimum number of PIM threads that keep the pipeline of the PIM core (i.e., UPMEM DPU) full~\cite{devaux2019, gomezluna2021benchmarking}. For this PIM core, a workload with performance saturation at 11 PIM threads can be considered a compute-bound workload on this PIM architecture, since the latency of instructions executed in the pipeline hides the latency of memory accesses~\cite{gomezluna2021benchmarking}. 

Second, using \fixedp representation instead of \floatp (i.e., \texttt{LIN-INT32} instead of \texttt{LIN-FP32}) reduces the kernel time by an order of magnitude. The PIM cores used in our evaluation do \emph{not} \juang{natively} support \floatp arithmetic. Thus, \floatp operations are emulated, since the PIM cores only have integer arithmetic units~\cite{devaux2019, gomezluna2021benchmarking}. 

Third, \texttt{LIN-HYB} accelerates the PIM kernel by 41\% over \texttt{LIN-INT32}. The speedup comes from the use of 8-bit integer multiplication, instead of the emulated 32-bit integer multiplication. 

Fourth, \texttt{LIN-BUI} achieves an additional 25\% speedup over \texttt{LIN-HYB} due to our custom multiplication operation (see Listing~\ref{lst:custom-mul}).
These results demonstrate that applying quantization on the training dataset can greatly increase the performance of PIM implementations without sacrificing much accuracy (see Section~\ref{sec:metrics}).

\subsubsection{\juang{Logistic Regression (LOG)}}
\label{sec:analysis_log}
\begin{figure}[h]
\centering
\includegraphics[width=1.0\linewidth]{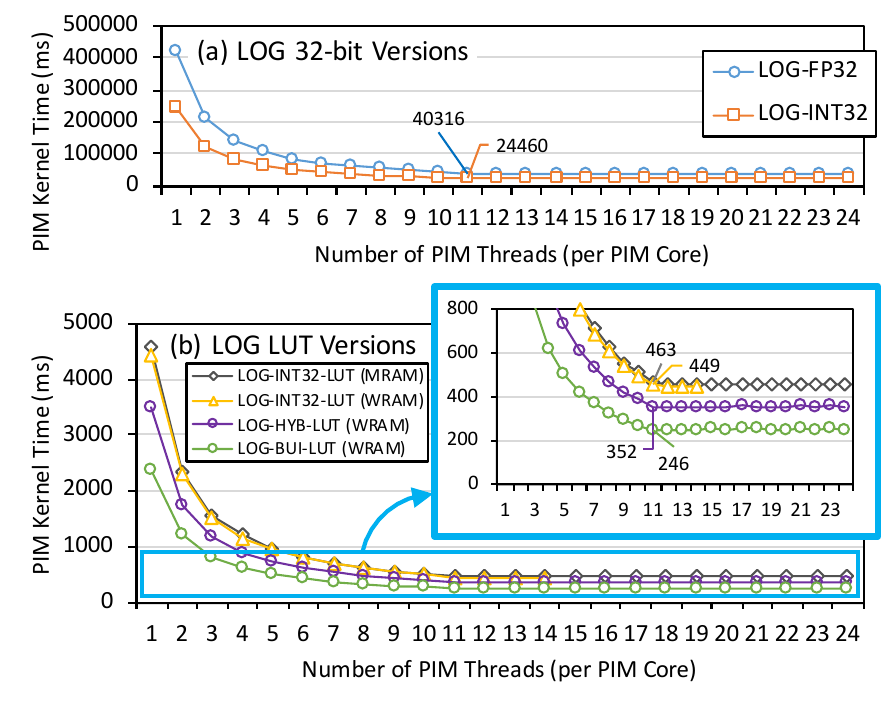}
\caption{Execution time (ms) of six versions of logistic regression \juang{using} 1-24 PIM threads in 1 PIM core.}
\label{fig:log-1dpu}
\end{figure}

Figure~\ref{fig:log-1dpu} shows the PIM kernel time of our versions of \texttt{LOG}. 
Figure~\ref{fig:log-1dpu}(a) shows the results for the two versions (\texttt{LOG-FP32}, \texttt{LOG-INT32}) that estimate sigmoid based on Taylor series. 
Although the 32-bit integer version reduces the kernel time by 65\% with respect to the 32-bit \floatp version, which uses emulated \floatp operations, the kernel time of both versions is very high due to the use of Taylor series, \juang{which require multiple iterations to achieve the necessary precision}. 
Figure~\ref{fig:log-1dpu}(b) shows the PIM kernel time of the LUT-based versions. 
We make five observations. 
First, the performance of all \texttt{LOG} versions saturates at 11 PIM threads, for the same reason as \texttt{LIN} versions. 

Second, \texttt{LOG-INT32-LUT (MRAM)} results in a speedup of $53\times$ over \texttt{LOG-INT32}. This demonstrates the benefit 
\juang{of converting computation to memory accesses using LUTs} in PIM architectures. 

Third, there is very little speedup (3\%) coming from placing the LUT in the scratchpad (WRAM of UPMEM DPUs). The LUT query is just one memory access and its cost is negligible compared to the rest of computation. 

Fourth, the use of 8-bit integer multiplication allows \texttt{LOG-HYB-LUT (WRAM)} to outperform \texttt{LOG-INT32-LUT (WRAM)} by 28\%. 

Fifth, the custom multiplication used by \texttt{LOG-BUI-LUT (WRAM)} provides an extra 43\% speedup over \texttt{LOG-HYB-LUT (WRAM)}.

\subsubsection{\juang{Decision Tree (DTR)}}
\begin{figure}[h]
\centering
\includegraphics[width=1.0\linewidth]{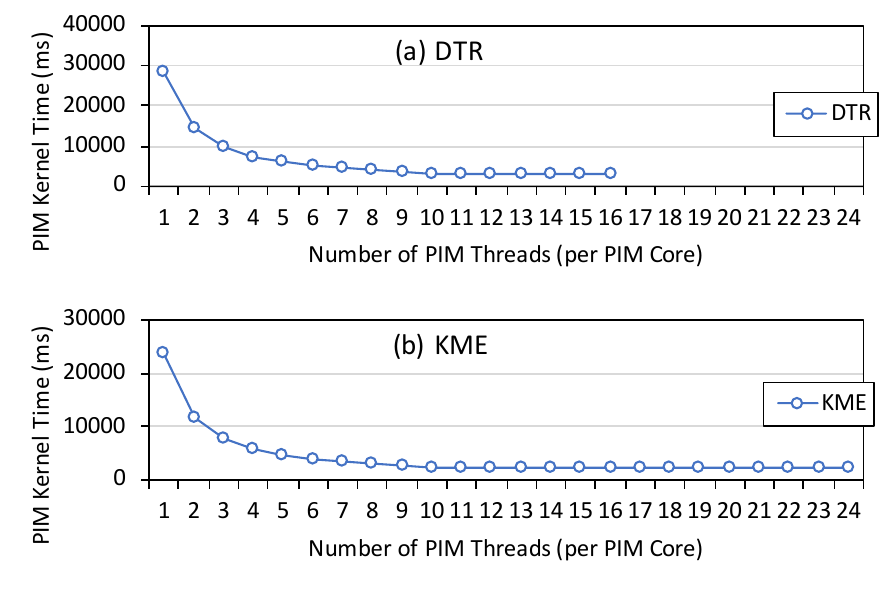}
\caption{Execution time (ms) of decision tree (a) \juang{using} 1-16 PIM threads in 1 PIM core and \km clustering (b) \juang{using} 1-24 PIM threads in 1 PIM core.}
\label{fig:dtr-kme-1dpu}
\end{figure}

Figure~\ref{fig:dtr-kme-1dpu}(a) shows the PIM kernel time of \texttt{DTR}. 
We make \juang{three} observations. 
First, the performance of \texttt{DTR} saturates at 11 PIM threads, for the same reason as \texttt{LIN} versions. 

\juang{Second, the optimized data layout of \texttt{DTR} (Section~\ref{sec:dtree}) ensures that data is accessed at maximum bandwidth and, thus, the pipeline latency hides the latency of memory accesses.}

Third, the maximum possible number of PIM threads is 16. This is due to the usage of the local scratchpad memory in the PIM core. The amount of memory needed by each PIM thread limits the maximum number of PIM threads to 16.

\subsubsection{\juang{\km Clustering (KME)}}
Figure~\ref{fig:dtr-kme-1dpu}(b) shows the PIM kernel time of \texttt{KME}. 
The performance of \texttt{KME} saturates at 11 PIM threads, for the same reason as \texttt{LIN} versions. 

In summary, the four workloads saturate at 11 PIM threads, as it is expected in the PIM cores we use in our experiments (i.e., UPMEM DPUs) for workloads where the pipeline latency hides the memory access latency~\cite{gomezluna2021benchmarking}. As a result, these workloads behave as compute-bound on these PIM cores (with high memory bandwidth but relatively slow pipeline), as opposed to their memory-bound behavior on processor-centric systems (see Section~\ref{sec:background}).

\subsection{Performance Scaling}
\label{sec:scaling}
We evaluate performance scaling characteristics of our ML workloads using weak scaling and strong scaling experiments. 
For weak scaling (Section~\ref{sec:weak}), we run experiments on 1 rank (from 1 to 64 PIM cores). Our goal is to evaluate how the performance scales with the number of PIM cores 
for a fixed problem size per processing element.
For strong scaling (Section~\ref{sec:strong}), we run experiments on 32 ranks (from 256 to 2,048 PIM cores). Our goal is to evaluate how the performance of our ML workloads scales with the number of PIM cores 
for a fixed problem size.

\subsubsection{Weak Scaling}
\label{sec:weak}

\begin{figure*}[h]
\centering
\includegraphics[width=1.0\linewidth]{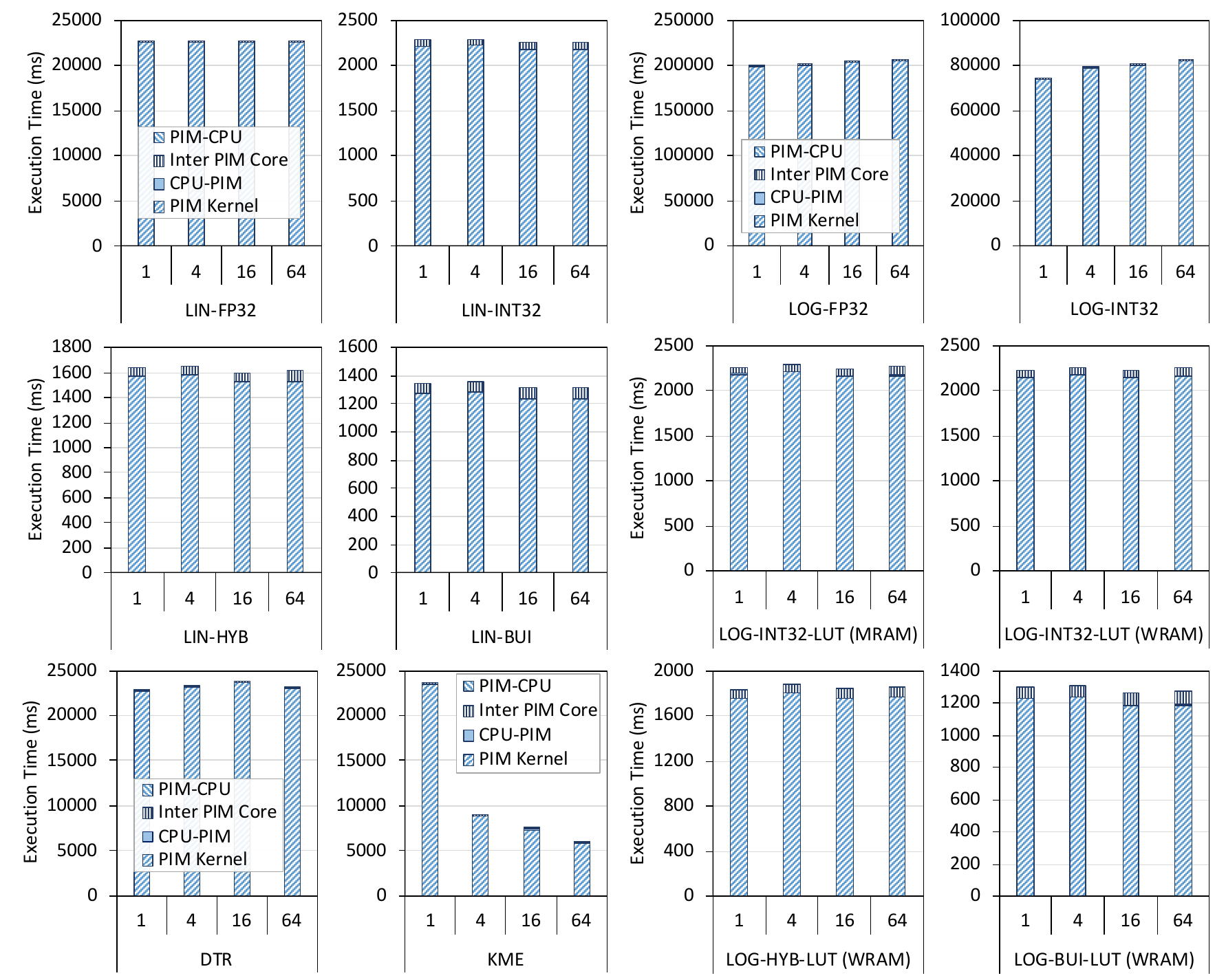}
\caption{Execution time (ms) of ML workloads on 1, 4, 16, and 64 PIM cores using weak scaling. Inside a PIM core, we use the best performing number of PIM threads (Section~\ref{sec:analysis}).}
\label{fig:weak}
\end{figure*}

Figure~\ref{fig:weak} shows weak scaling results on 1-64 PIM cores for all versions of our ML workloads. 
Each bar presents the total execution time broken down into (1) execution time of the PIM kernel (i.e., PIM Kernel), communication time between the host CPU and the PIM cores (i.e., CPU-PIM and PIM-CPU times), and communication time between PIM cores (i.e., Inter PIM Core). 
We make the following observations from the figure. 

First, we observe linear scaling of the PIM kernel time of all \texttt{LIN} versions, all \texttt{LOG} versions, and \texttt{DTR}. 
However, the PIM kernel time of \texttt{KME} reduces as we increase the number of PIM cores.  
This is caused by the fact that the \km algorithm on average converges with fewer iterations on a larger dataset. The PIM kernel time per iteration does scale linearly.

Second, the fraction of total execution time spent on communication between the host CPU and the PIM cores (i.e., CPU-PIM and PIM-CPU\footnote{\texttt{DTR} and \texttt{KME} do \emph{not} need final PIM-CPU transfer. For \texttt{DTR}, the reason is that the tree is built iteratively on the host side, and the algorithm ends when the CPU declares termination on the tree build. For \texttt{KME}, the CPU is in charge of the final cluster assignment once convergence has been declared.} 
times) and between PIM cores (i.e., Inter PIM Core) is negligible compared to the PIM kernel time for all versions. 
For all \texttt{LIN} versions, all \texttt{LOG} versions, \texttt{DTR}, and \texttt{KME}, the sum of CPU-PIM, Inter PIM Core, and PIM-CPU times takes less than 7\% of the total execution time. 

\subsubsection{Strong Scaling}
\label{sec:strong}

\begin{figure*}[h]
\centering
\includegraphics[width=1.0\linewidth]{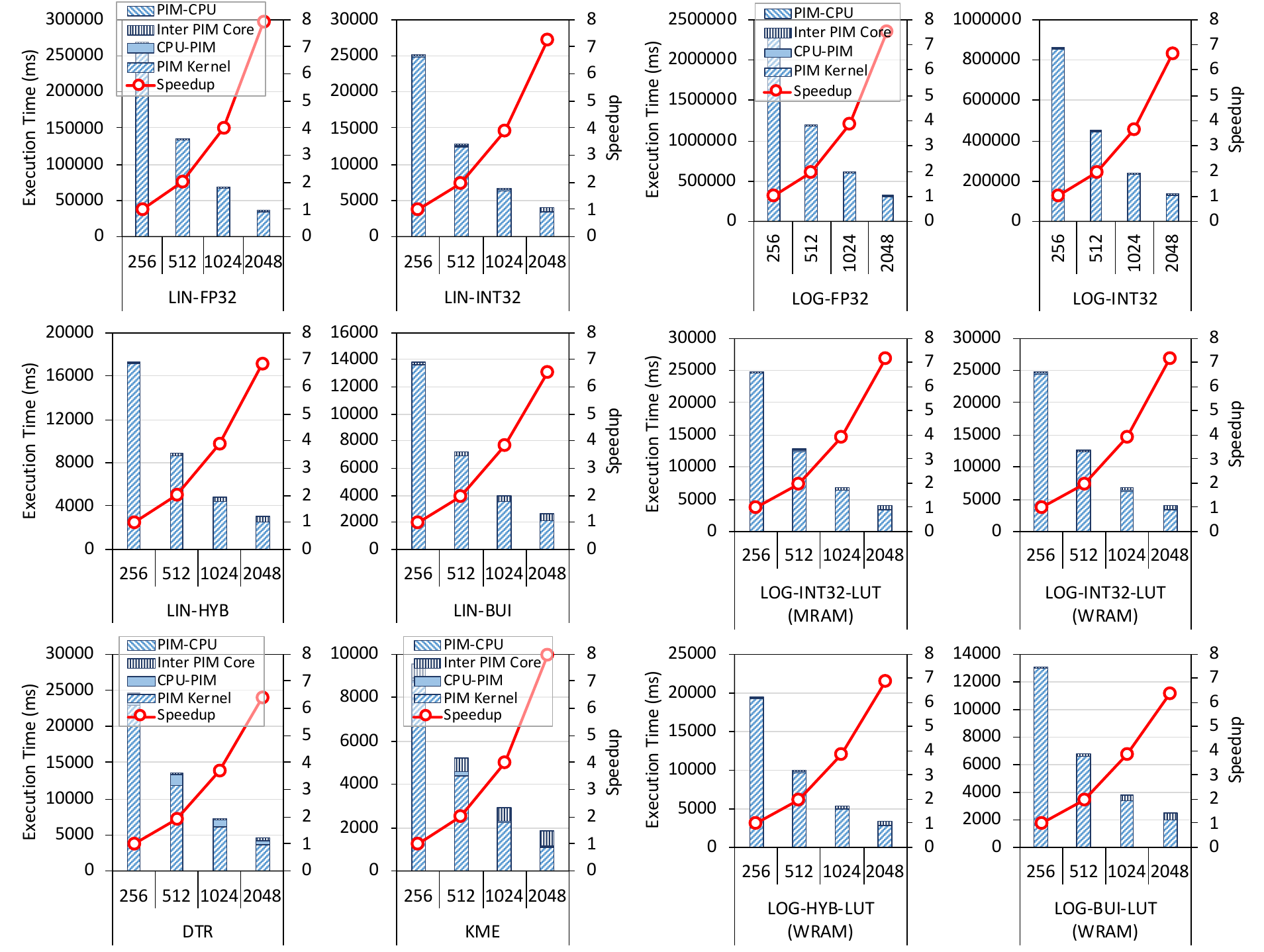}
\caption{Execution time (ms) of ML workloads on 256, 512, 1,024, and 2,048 PIM cores using strong scaling (left y-axis), and speedup of the PIM kernel normalized to the performance of 256 PIM cores (right y-axis). Inside a PIM core, we use the best performing number of PIM threads (Section~\ref{sec:analysis}).}
\label{fig:strong}
\end{figure*}

Figure~\ref{fig:strong} shows strong scaling results on 256-2,048 PIM cores for all versions of our ML workloads. 
Each bar (left y-axis) presents the total execution time broken down into (1) execution time of the PIM kernel (i.e., PIM Kernel), communication time between the host CPU and the PIM cores (i.e., CPU-PIM and PIM-CPU times), and communication time between PIM cores (i.e., Inter PIM Core). Each red line (right y-axis) represents the speedup of a PIM kernel normalized to the performance of 256 PIM cores. 
We make the following observations. 

First, we observe that the PIM kernel time scales linearly with the number of PIM cores. 
The speedup of 2,048 PIM cores over 256 PIM cores is between $6.37\times$ and $7.98\times$. 

Second, the overhead of communication between PIM cores (i.e., Inter PIM Core) is tolerable for all ML workloads. The largest fraction of Inter PIM Core over the total execution time is 36\% for \texttt{KME} with 2,048 PIM cores. Even so, 2,048 PIM cores provide the lowest total execution time of \texttt{KME}.

Third, the communication time between the host CPU and the PIM cores (i.e., CPU-PIM and PIM-CPU times) represents a negligible fraction of the total execution time of all ML workloads.

\subsection{Comparison to CPU and GPU}
\label{sec:cpugpu}
We compare our implementations of ML workloads on a PIM system to state-of-the-art CPU and GPU implementations of the same workloads in terms of performance and quality. 
Table~\ref{tab:comparison} indicates the sources of these CPU and GPU implementations. 

\begin{table}[h]
\begin{center}
\caption{CPU and GPU implementations of ML workloads.}
\label{tab:comparison}
\resizebox{1.0\linewidth}{!}{
\begin{tabular}{|l||l|l|}
\hline
\textbf{ML workload} & \textbf{CPU implementation} & \textbf{GPU implementation} \\
\hline
\hline
Linear regression & Intel MKL~\cite{mkl} & NVIDIA cuBLAS~\cite{cublas} \\ \hline
Logistic regression & Intel MKL~\cite{mkl} & NVIDIA cuBLAS~\cite{cublas} \\ \hline
Decision tree & Scikit-learn~\cite{pedregosa2011scikit} & RAPIDS~\cite{rapids} \\ \hline
Kmeans & Scikit-learn~\cite{pedregosa2011scikit} & RAPIDS~\cite{rapids} \\ \hline
\end{tabular}

}
\end{center}
\end{table}

Our goal is to evaluate the potential of a general-purpose PIM system (Table~\ref{tab:pim-cpugpu}) for acceleration of ML workloads. 
We use an Intel Xeon Silver 4215 CPU~\cite{xeon-4215} and an NVIDIA A100 GPU~\cite{a100} based on the Ampere architecture~\cite{ampere} as baseline processor-centric architectures. 
Table~\ref{tab:pim-cpugpu} summarizes their key characteristics.

For the PIM system performance measurements, we include the time spent in the PIM cores ("PIM Kernel"), the time spent for inter-PIM-core synchronization ("Inter PIM"), and the time spent in the initial CPU-PIM and the final PIM-CPU transfers ("CPU-PIM", "PIM-CPU"). 
For the GPU performance measurements, we include the kernel time ("GPU Kernel"), and the initial CPU-GPU and the final GPU-CPU transfer times ("CPU-GPU", "GPU-CPU"). 
The results that we show in this section correspond to the best configurations in terms of CPU threads (for the CPU versions), GPU threads per block and thread blocks (for the GPU versions), and PIM cores and PIM threads (for the PIM versions). 
\juang{We open-source all configurations for reproducibility~\cite{gomezluna2023repo}.}

\subsubsection{\juang{Linear Regression (LIN)}}
\begin{figure*}[h]
\centering
\includegraphics[width=0.7\linewidth]{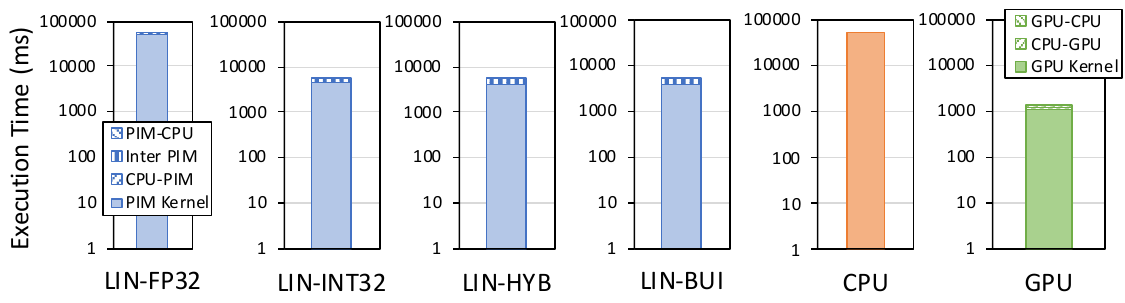}
\caption{Execution time (ms) of \texttt{LIN} on PIM, CPU, and GPU with the SUSY dataset. For the PIM versions (\texttt{LIN-*}), we use the best performing number of PIM cores (2,524 for all four versions). Inside a PIM core, we use the best performing number of PIM threads (Section~\ref{sec:analysis}).} 
\label{fig:susy-q}
\end{figure*}

Figure~\ref{fig:susy-q} shows the execution times of \texttt{LIN} versions on PIM, CPU, and GPU with the SUSY dataset~\cite{susy}. 
We apply symmetric quantization~\cite{zmora2021quantization, gholamisurvey} to this dataset, in order to be able to evaluate our integer versions. 
We make the following observations. 
First, \texttt{LIN-FP32} is heavily burdened by the use of \floatp arithmetic, which is not natively supported by the PIM system we use in our evaluation (i.e., UPMEM-based PIM system)~\cite{gomezluna2021benchmarking}. Despite that, \texttt{LIN-FP32} is 13\% faster than the CPU version. 
Second, \texttt{LIN-INT32} is $8.5\times$ faster than \texttt{LIN-FP32}. This is the result of using natively supported instructions (even though 32-bit integer multiplication is emulated in the UPMEM PIM architecture)~\cite{gomezluna2021benchmarking}. 
Third, \texttt{LIN-HYB} and \texttt{LIN-BUI} further improve the performance. The kernel time of \texttt{LIN-HYB} is 10\% lower than that of \texttt{LIN-INT32} due to the use of hybrid precision. Our custom multiplication in \texttt{LIN-BUI} reduces the kernel time by an additional 4\%. 
Fourth, the GPU version is $4.1\times$ faster than our \texttt{LIN-BUI}, since the A100 (1) has much higher compute throughput than the PIM system that we use in our experiments, and (2) its memory bandwidth is only $39\%$ lower than the bandwidth of the PIM system (Table~\ref{tab:pim-cpugpu}).

\subsubsection{\juang{Logistic Regression (LOG)}}
\begin{figure}[h]
\includegraphics[width=1.0\linewidth]{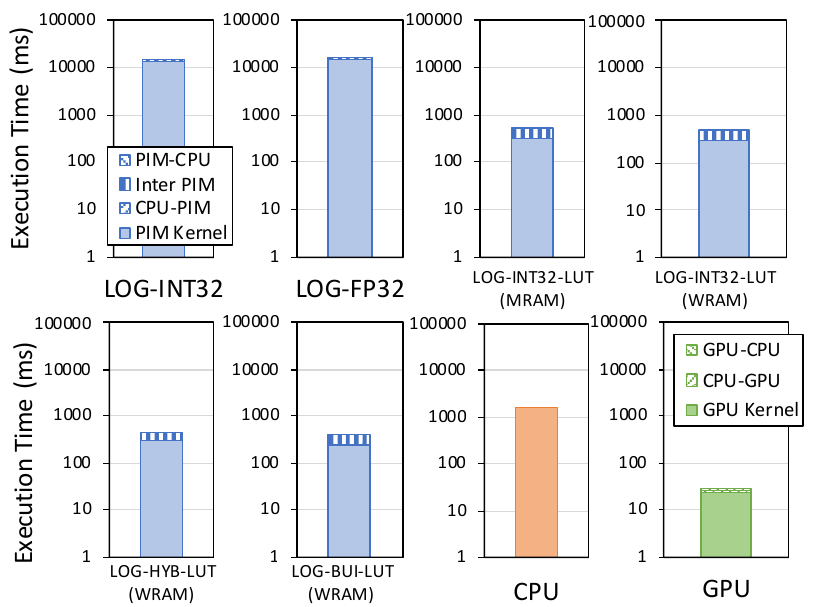}
\caption{Execution time (ms) of \texttt{LOG} on PIM, CPU, and GPU with the Skin segmentation dataset. For the PIM versions (\texttt{LOG-*}), we use the best performing number of PIM cores (2,524 for \texttt{LOG-FP32} and \texttt{LOG-INT32}; 320 for \texttt{LOG-INT32-LUT (MRAM)} and \texttt{LOG-INT32-LUT (WRAM)}; 256 for \texttt{LOG-HYB-LUT (WRAM)} and \texttt{LOG-BUI-LUT (WRAM)}). Inside a PIM core, we use the best performing number of PIM threads (Section~\ref{sec:analysis}).}
\label{fig:skin}
\end{figure}

Figure~\ref{fig:skin} shows the execution times of \texttt{LOG} versions on PIM, CPU, and GPU with the Skin segmentation dataset~\cite{skin}. 
We make four observations from these results. 
First, \texttt{LOG-FP32} and \texttt{LOG-INT32} PIM versions are almost $10\times$ slower than the CPU version. The reason is the high cost of sigmoid estimation with Taylor series \juang{due to their iterative nature (as mentioned in Section~\ref{sec:analysis_log})}. 
Second, \texttt{LOG-INT32} is $17\%$ faster than \texttt{LOG-FP32} due to the faster integer arithmetic~\cite{gomezluna2021benchmarking}. 
Third, \juang{replacing Taylor series with} the use of LUTs (in \texttt{LOG-INT32-LUT (MRAM)}, \texttt{LOG-INT32-LUT (WRAM)}, \texttt{LOG-HYB-LUT (WRAM)}, and \texttt{LOG-BUI-LUT (WRAM)}) to estimate sigmoid accelerates the PIM versions by almost two orders of magnitude. For example, \texttt{LOG-INT32-LUT (WRAM)} is $3.3\times$ and \texttt{LOG-BUI-LUT (WRAM)} is $3.9\times$ faster than the CPU version. 
Fourth, even though the GPU version is significantly faster than all PIM versions (e.g., $16.5\times$ faster than \texttt{LOG-BUI-LUT (WRAM)}), the gap between \juang{GPU and PIM} is greatly reduced by using appropriate optimizations in PIM codes (e.g., LUTs, custom multiplication).

Table~\ref{tab:linlog-error} shows the training error rate (\%) of all versions of \texttt{LIN} (with the SUSY dataset) and \texttt{LOG} (with the Skin segmentation dataset). We make several observations. 
First, the training error rates of the 32-bit \floatp versions (i.e., \texttt{LIN-FP32}, \texttt{LOG-FP32}) is the same as that of the CPU and the GPU versions. 
Second, the training error rates of the PIM versions of \texttt{LIN} and \texttt{LOG} that use quantized datasets are greater than those of the CPU and GPU versions, but they may still be acceptable (i.e., $<20\%$ for \texttt{LIN} and $<9\%$ for \texttt{LOG}) for some applications~\cite{wijaya2016hybrid, chang2008partitioned, dedeturk2020spam, sivasankari2022detection, martin2022iot, akbar2017predictive, sarangdhar2017machine}.

\begin{table}[h]
\begin{center}
\caption{Training error rate (\%) of \texttt{LIN} and \texttt{LOG} versions on PIM, CPU, and GPU.}
\label{tab:linlog-error}
\resizebox{1.0\linewidth}{!}{
\begin{tabular}{|l|l||c|}
\hline
\textbf{ML workload} & \textbf{Version} & \textbf{Training error rate (\%)} \\
\hline
\hline
\multirow{6}{*}{Linear regression} & \texttt{LIN-FP32} & 13.88 \\ \cline{2-3}
& \texttt{LIN-INT32} & 18.68 \\ \cline{2-3}
& \texttt{LIN-HYB} & 18.68 \\ \cline{2-3}
& \texttt{LIN-BUI} & 18.68 \\ \cline{2-3}
& CPU~\cite{mkl} & 13.88 \\ \cline{2-3}
& GPU~\cite{cublas} & 13.88 \\ \hline
\multirow{8}{*}{Logistic regression} & \texttt{LOG-FP32} & 7.58 \\ \cline{2-3}
& \texttt{LOG-INT32} & 8.72 \\ \cline{2-3}
& \texttt{LOG-INT32-LUT (MRAM)} & 8.72 \\ \cline{2-3}
& \texttt{LOG-INT32-LUT (WRAM)} & 8.98 \\ \cline{2-3}
& \texttt{LOG-HYB-LUT (WRAM)} & 8.98 \\ \cline{2-3}
& \texttt{LOG-BUI-LUT (WRAM)} & 8.98 \\ \cline{2-3}
& CPU~\cite{mkl} & 7.58 \\ \cline{2-3}
& GPU~\cite{cublas} & 7.58 \\ \hline
\end{tabular}

}
\end{center}
\end{table}

\subsubsection{\juang{Decision Tree (DTR)}}
\begin{figure*}[h]
\centering
\includegraphics[width=0.7\linewidth]{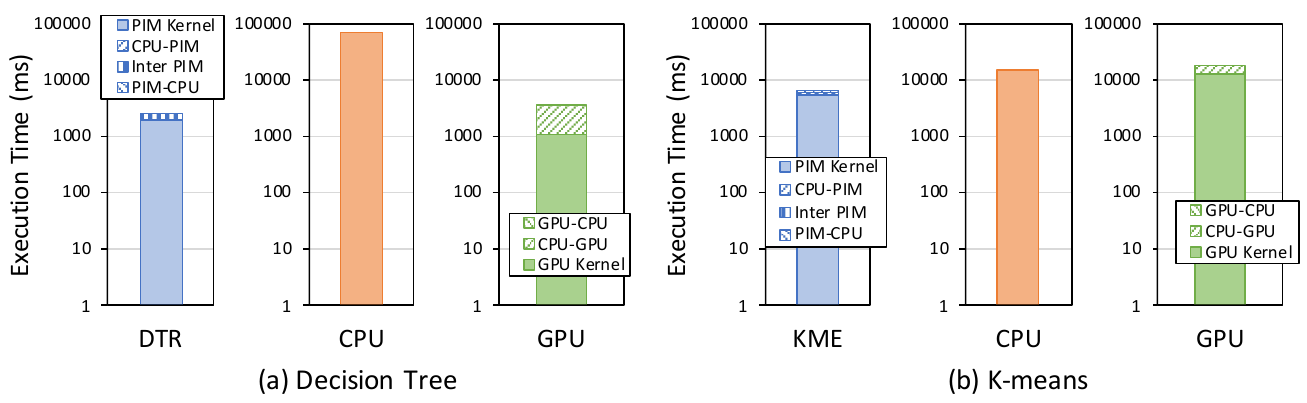}
\caption{Execution time (ms) of \texttt{DTR} (a) and \texttt{KME} (b) on PIM, CPU, and GPU with the Higgs boson segmentation dataset. For the PIM versions (\texttt{DTR}, \texttt{KME}), we use the best performing number of PIM cores (1,024 for \texttt{DTR}; 2,524 for \texttt{KME}). Inside a PIM core, we use the best performing number of PIM threads (Section~\ref{sec:analysis}).}
\label{fig:boson}
\end{figure*}

Figure~\ref{fig:boson}(a) shows the execution times of \texttt{DTR} versions on PIM, CPU, and GPU with the Higgs boson dataset~\cite{Dua:2019}. 
We make \juang{two} observations. 
First, the PIM version of \texttt{DTR} outperforms the CPU version and the GPU version by $27\times$ and $1.34\times$, respectively. Since \texttt{DTR} mostly uses \juang{comparison operations (e.g., comparing a feature value to a threshold)}, the PIM version can take advantage of the large internal bandwidth of the PIM system without being burdened by other costly arithmetic operations. 
Second, $70\%$ of the execution time of the GPU version of \texttt{DTR} is spent on moving data between the host CPU and the GPU, while only $27\%$ of the execution time of the PIM version is due to communication between the host CPU and the PIM cores or between PIM cores. 
\juang{The fact that} the host CPU and the PIM cores \juang{are connected} through memory channels is an advantage over the GPU, which uses PCIe bus, as the memory channels provide higher bandwidth. 

\begin{figure*}[h]
\centering
\includegraphics[width=0.7\linewidth]{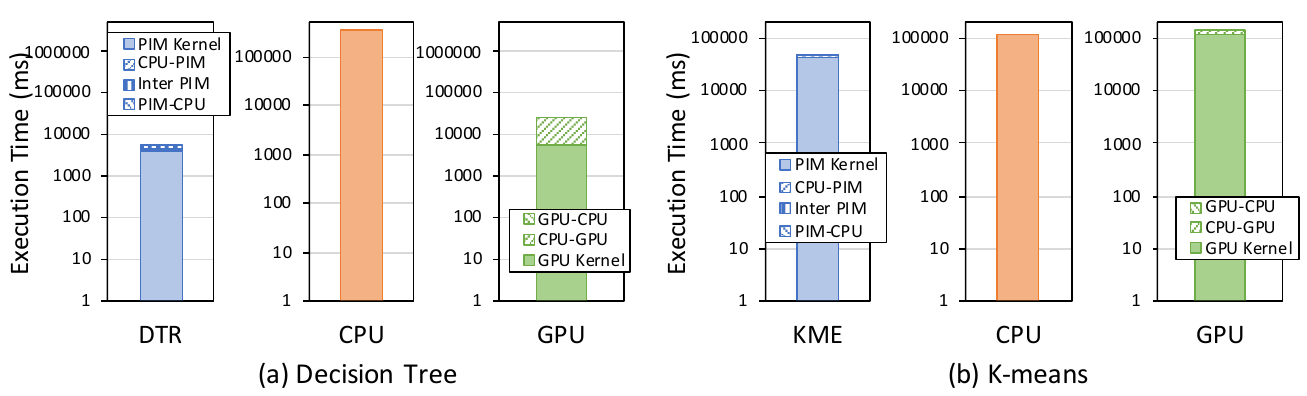}
\caption{\juangg{Execution time (ms) of \texttt{DTR} (a) and \texttt{KME} (b) on PIM, CPU, and GPU with the Criteo dataset (partial dataset). For the PIM versions (\texttt{DTR}, \texttt{KME}), we use the best performing number of PIM cores (2,524 for both \texttt{DTR} and \texttt{KME}). Inside a PIM core, we use the best performing number of PIM threads (Section~\ref{sec:analysis}).}}
\label{fig:criteo-quarter}
\end{figure*}

\juangg{For \texttt{DTR}, we also run experiments using the Criteo dataset~\cite{criteo}. 
Figure~\ref{fig:criteo-quarter}(a) shows the execution times of \texttt{DTR} versions on PIM, CPU, and GPU with a quarter of day 0. 
We make two observations in line with the observations from the results with the Higgs boson dataset (Figure~\ref{fig:boson}(a)). 
First, the speedup of the PIM version of \texttt{DTR} over the CPU version and the GPU version increases to $62\times$ and $4.5\times$, respectively. Since the part of the Criteo dataset that we use is significantly larger than the Higgs boson dataset, there is enough work to keep all 2,524 PIM cores busy (the best performing number of PIM cores for the Higgs boson dataset is 1,024). This explains the increased speedups with respect to those for the Higgs boson dataset. 
Second, the percentage of the execution time devoted to communication remains $27\%$ for the PIM version, while it increases to $77\%$ for the GPU version. This also explains the increased speedup of the PIM version over the GPU version.}

\begin{figure}[h]
\centering
\includegraphics[width=1.0\linewidth]{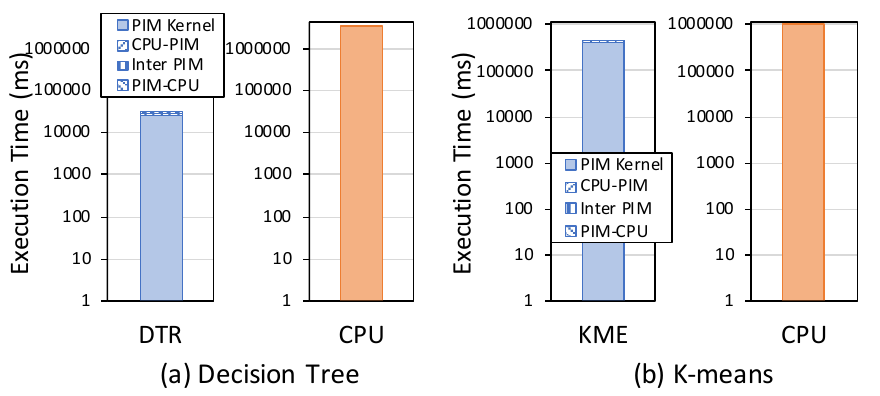}
\caption{\juangg{Execution time (ms) of \texttt{DTR} (a) and \texttt{KME} (b) on PIM and CPU with the Criteo dataset (2 days). For the PIM versions (\texttt{DTR}, \texttt{KME}), we use the best performing number of PIM cores (2,524 for both \texttt{DTR} and \texttt{KME}). Inside a PIM core, we use the best performing number of PIM threads (Section~\ref{sec:analysis}).}}
\label{fig:criteo-2days}
\end{figure}

\juangg{Figure~\ref{fig:criteo-2days}(a) shows the execution times of \texttt{DTR} versions on PIM and CPU with two days (days 0 and 1) of the Criteo dataset~\cite{criteo}. 
A main observation is that the PIM version outperforms the CPU version by $113\times$. A possible explanation for this high speedup is that the large dataset size causes many page faults on the CPU.} 

\juanggr{Using the Criteo dataset~\cite{criteo}, we also compare the PIM version of \texttt{DTR} to another, more optimized, CPU baseline, the Intel extension for Scikit-learn~\cite{intel-scikit, intel-scikit-repo}. 
For a quarter of day 0, the PIM version is $21\times$ faster than Intel's CPU version. 
For days 0 and 1, the PIM version is $36\times$ faster than Intel's CPU version. 
Despite the Intel extension for Scikit-learn~\cite{intel-scikit, intel-scikit-repo} is significantly faster than the original Scikit-learn~\cite{pedregosa2011scikit}, the speedups that our PIM version of \texttt{DTR} provides are still considerable.}

Table~\ref{tab:dtr-error} shows the training accuracy of \texttt{DTR} versions on PIM, CPU, and GPU \juangg{for the Higgs boson dataset}. 
We observe that the accuracy of our PIM version (0.65635) is very similar to the accuracy of the CPU version (0.65581), and only slightly smaller than that of the GPU version (0.70462).

\begin{table}[h]
\begin{center}
\caption{Training accuracy of \texttt{DTR} versions on PIM, CPU, and GPU.}
\label{tab:dtr-error}
\resizebox{0.75\linewidth}{!}{
\begin{tabular}{|l|l||c|}
\hline
\textbf{ML workload} & \textbf{Version} & \textbf{Training accuracy} \\
\hline
\hline
\multirow{3}{*}{Decision tree} & \texttt{DTR} (PIM) & 0.65635 \\ \cline{2-3}
& CPU~\cite{pedregosa2011scikit} & 0.65581 \\ \cline{2-3}
& GPU~\cite{rapids} & 0.70462 \\ \hline
\end{tabular}

}
\end{center}
\end{table}

\subsubsection{\juang{\km Clustering (KME)}}
\juang{Figure~\ref{fig:boson}(b) shows the execution times of \texttt{KME} versions on PIM, CPU, and GPU with the Higgs boson dataset~\cite{Dua:2019}. 
We observe that the PIM version of \texttt{KME} is $2.8\times$ faster than the CPU version and $3.2\times$ faster than the GPU version. Similar to \texttt{DTR}, \texttt{KME} does \emph{not} use costly arithmetic operations but mainly 16-bit integer arithmetic.}

\juangg{For \texttt{KME}, we also run experiments using the Criteo dataset~\cite{criteo}. 
Figure~\ref{fig:criteo-quarter}(b) shows the execution times of \texttt{DTR} versions on PIM, CPU, and GPU with a quarter of day 0. 
Figure~\ref{fig:criteo-2days}(b) shows the execution times with two days (days 0 and 1) of the Criteo dataset. 
We observe that the speedups of the PIM version over the CPU version \juanggr{($2.7\times$ with a quarter of day 0, and $2.4\times$ with two days)} and the GPU version \juanggr{($3.2\times$ with a quarter of day 0)} remain similar to the speedups with the Higgs boson dataset (Figure~\ref{fig:boson}(b)). 
This can be explained by the fact that the best performing number of PIM cores is the maximum (2,524) for both datasets. Thus, for \texttt{KME}, a larger dataset does not increase the speedup of the PIM version over the CPU and the GPU versions.}

\juanggr{We also compare the PIM version of \texttt{KME} to the Intel extension for Scikit-learn~\cite{intel-scikit, intel-scikit-repo} using the Criteo dataset~\cite{criteo}. 
The Intel extension for Scikit-learn~\cite{intel-scikit, intel-scikit-repo} is almost twice as fast as the original Scikit-learn~\cite{pedregosa2011scikit}, but our PIM version of \texttt{KME} is still faster. 
For a quarter of day 0, the PIM version is $1.42\times$ faster than Intel's CPU version. 
For days 0 and 1, the PIM version is $1.37\times$ faster than Intel's CPU version.}

Table~\ref{tab:kme-error} shows the similarity of the clusterings (given by the adjusted Rankd index) produced by \texttt{KME} versions on PIM, CPU, and GPU \juangg{for the Higgs boson dataset}. 
The adjusted Rand index between the PIM version and the CPU version is 0.999985, while the adjusted Rand index between the GPU version and the CPU version is significantly lower (0.758579).

\begin{table}[h]
\begin{center}
\caption{Adjusted Rand index of \texttt{KME} versions on PIM, CPU, and GPU.}
\label{tab:kme-error}
\resizebox{0.8\linewidth}{!}{
\begin{tabular}{|l|l||c|}
\hline
\textbf{ML workload} & \textbf{Version} & \textbf{Adjusted Rand index} \\
\hline
\hline
\multirow{3}{*}{K-means} & \texttt{KME} (PIM) & 0.999985 \\ \cline{2-3}
& CPU~\cite{pedregosa2011scikit} & 1 \\ \cline{2-3}
& GPU~\cite{rapids} & 0.758579 \\ \hline
\end{tabular}

}
\end{center}
\end{table}

\ignore{
We use Intel RAPL~\cite{rapl} on the CPU and NVIDIA SMI~\cite{smi} on the GPU for energy measurements. 
We obtain the energy consumed by the PIM system with an analytical model provided by the UPMEM company~\cite{upmem-sdk, upmem}, which is based on the number of execution cycles, executed instructions, memory accesses (WRAM and MRAM) in all DPUs (i.e., processing elements). The results consider only the energy of the PIM chips.

\jgl{Evaluation in terms of TCO (besides performance and energy).}

\todo{We complete the paper before obtaining the energy results. 
We need energy results for a strong submission, but the first version may not have them.}
}

\section{Key Takeaways and Recommendations}
\label{sec:discussion}

In this section, we summarize our key observations, takeaways, and recommendations that stem from our analysis of four ML training workloads on a state-of-the-art general-purpose PIM architecture. 
There are four subsections that are titled after four widely-accepted facts about near-bank PIM architectures. 
For each of them, we first state general key observations that are derived from the particular fact. Then, we provide insights derived from our work in the form of key takeaways and recommendations.

\subsection{PIM Cores Are Compute-bound}
PIM cores~\cite{devaux2019, lee2021hardware, kwon202125, lee2022isscc} are wimpy processors (operating at relatively low frequency) with high memory bandwidth, especially for streaming memory access patterns~\cite{gomezluna2021benchmarking}. 
As a result, for real-world workloads, the compute throughput tends to saturate much more frequently than the memory bandwidth, and the pipeline latency hides the memory access latency. 

\textbf{Key Takeaway \#1.} Even ML training workloads (e.g., linear regression, logarithmic regression, decision tree, \km) that are bound by memory access due to their low arithmetic intensity in processor-centric systems (e.g., CPU, GPU) behave as compute-bound when running on PIM cores.

\textbf{Recommendation \#1.} Maximize the utilization of PIM cores by keeping their pipeline fully busy. 
For example, in the UPMEM PIM architecture~\cite{devaux2019}, which has fine-grained multithreaded scalar cores, we recommend to schedule 11 or more PIM threads (Section~\ref{sec:analysis}), which is the minimum number of PIM threads to saturate the pipeline throughput. 
In SIMD-based PIM architectures~\cite{lee2021hardware, kwon202125, lee2022isscc}, a recommendation would be to maximize SIMD utilization by minimizing divergence across SIMD lanes~\cite{ddca.spring2020.simd}.

\subsection{PIM Cores Have Limited Instruction Sets}
PIM cores~\cite{devaux2019, lee2021hardware, kwon202125, lee2022isscc} have limited instruction sets. As such, they do \emph{not} support natively a large variety of arithmetic operations and datatypes. 
For example, the UPMEM PIM architecture~\cite{devaux2019} does \emph{not} support \floatp operations or 32-bit integer multiplication/division (only emulated by the runtime library)~\cite{gomezluna2021benchmarking}. 
AiM~\cite{lee2022isscc} and HBM-PIM~\cite{lee2021hardware, kwon202125} only support multiplication and addition of 16-bit \floatp values.

\textbf{Key Takeaway \#2.} Workloads that require arithmetic operations or datatypes that are not natively supported by PIM cores will either (1) run at low performance due to instruction emulation (e.g., \floatp operations in UPMEM PIM), or (2) cannot be implemented on those PIM architectures (e.g., workloads requiring operations other than multiplication and addition in AiM or HBM-PIM).

\textbf{Recommendation \#2.} ML workloads (e.g., \texttt{LIN}, \texttt{LOG}) can employ \fixedp representation if PIM cores do not support \floatp operations (e.g., UPMEM PIM) without sacrificing much accuracy (Section~\ref{sec:metrics}).

\textbf{Recommendation \#3.} Quantization can be used to take advantage of native hardware support, if PIM cores \juang{natively} support \juang{only} limited precision. For example, using hybrid precision after quantizing the training dataset can provide significant performance improvements in the UPMEM PIM architecture (Section~\ref{sec:analysis}).

\textbf{Recommendation \#4.} Programmers (or better compilers) can optimize code at low level to better leverage the available native instructions and hardware (e.g., 8-bit integer multiplication in UPMEM DPUs). For example, we show that our custom 16-bit and 32-bit integer multiplications (Listing~\ref{lst:custom-mul}) significantly improve performance over compiler-generated code for quantized training datasets (Section~\ref{sec:analysis}).

\textbf{Key Takeaway \#3.} Memory-bound ML workloads that require mainly operations natively supported by the PIM architecture (e.g., 32-bit integer addition/subtraction in UPMEM PIM), \juang{such as decision tree and \km clustering}, leverage the large PIM bandwidth, and perform better than their state-of-the-art CPU and GPU counterparts. 
For example, \texttt{DTR} only requires comparison and 32-bit integer addition. Our PIM version of \texttt{DTR} outperforms the CPU and GPU versions by $27\times$ and $1.34\times$, respectively. 
Our PIM implementation of \texttt{KME} employs mainly 16-bit integer arithmetic. As a result, it outperforms the CPU and GPU version by by $2.8\times$ and $3.2\times$, respectively.

\subsection{PIM Cores Have High Memory Bandwidth}
Near-bank PIM cores leverage high aggregated memory bandwidth and low latency memory access. As a result, memory accesses are typically cheaper than computation (in particular, 32- or 64-bit integer arithmetic, \floatp arithmetic, transcendental functions) for workloads that are memory bound in processor-centric systems due to the relatively low frequency and simple pipelines of PIM cores~\cite{gomezluna2021benchmarking}.

PIM cores exploit memory bandwidth better when they perform streaming memory accesses to the memory banks, especially if memory arrays have large row buffers (e.g., DDR4 memory in UPMEM PIM).

\textbf{Recommendation \#5.} Programmers can 
\juang{convert computation to memory accesses} in PIM architectures by keeping pre-calculated operation results (e.g., LUTs, memoization) in memory. For example, our use of LUTs for sigmoid calculation in \texttt{LOG} results in a speedup of $53\times$ (\texttt{LOG-INT32-LUT (MRAM)} over \texttt{LOG-INT32}, Section~\ref{sec:analysis}). 

\textbf{Recommendation \#6.} For data structures of more than one dimension, programmers can optimize the data layout in a way that memory accesses are in streaming, thus exploiting higher sustained bandwidth. We show in this paper one example of optimized data layout for \texttt{DTR} on the UPMEM PIM architecture (Section~\ref{sec:dtree}).

\subsection{PIM Throughput Scales with Memory Capacity}
Since near-bank PIM cores are directly attached to the memory arrays, their number scales at the same pace as the number of memory arrays, banks, and chips in the memory subsystem. Consequently, the overall PIM throughput scales linearly. 

Large PIM-enabled memory allows training datasets to remain in memory during the whole training process. This represents an inherent advantage over processor-centric systems (e.g., CPU, GPU) where the whole training dataset needs to be moved to the processor in every iteration of the training process.

\textbf{Key Takeaway \#4.} Memory-bound ML training workloads, which need large training datasets, benefit from large PIM-enabled memory with many PIM cores. Even if PIM cores need to communicate via the host processor (e.g., in UPMEM PIM), the amount of data movement needed for intermediate results is minimal with respect to the size of the whole training dataset. Our strong scaling and weak scaling characterization of four ML training workloads (Section~\ref{sec:scaling}) demonstrates that.


\section{Related Work}


To our knowledge, this is the first work that \emph{comprehensively} evaluates the benefits of a \emph{real} general-purpose \juang{processing-in-memory (PIM) system} for ML training \juang{workloads}. We briefly summarize prior works on PIM acceleration of Deep Learning (DL) and other ML algorithms.

\textbf{PIM for DL inference.} \juang{Many} prior works focus on accelerating DL inference using different PIM solutions. This includes both proposals from Academia~\cite{gao2017tetris,boroumand.asplos18,azarkhish2017neurostream,kwon2019tensordimm,deng.dac2018,ke2020recnmp,shin2018mcdram,boroumand2021google,cordeiro2021machine,lee2021task,park2021high,park2021trim,cho2020mcdram,kim2020mvid} and Industry~\cite{lee2022isscc,kwon202125,lee2021hardware,ke2021near,niu2022isscc}, targeting various types of DL models, including convolutional neural networks~\cite{gao2017tetris,kwon202125,lee2021hardware,boroumand.asplos18,azarkhish2017neurostream,deng.dac2018,ke2020recnmp,shin2018mcdram,boroumand2021google,cordeiro2021machine,lee2021task,park2021high,cho2020mcdram}, recurrent neural networks~\cite{lee2022isscc,boroumand2021google,kim2020mvid}, and recommendation systems~\cite{ke2021near,niu2022isscc,kwon2019tensordimm,park2021trim}. Our work differs from such works since we focus on classic ML algorithms (i.e., regression, classification, clustering) using a real-world general-purpose PIM architecture (i.e., the commercially-available UPMEM PIM architecture~\cite{upmem}).

\textbf{PIM for DL training.} Another body of works leverages PIM techniques to accelerate DL training~\cite{luo2020benchmark,sun2020energy,luo2020accelerating,li20203d,su202015,imani2019floatpim,jiang2019cimat,cheng2018time,liu2018processing,marinella2018multiscale,schuiki2018scalable,hasan2017fast,gu2020dlux}. These works mainly utilize the analog computation capabilities \juang{(e.g., for matrix vector multiplication)} of non-volatile memory (NVM) technologies to implement training of deep neural networks~\cite{luo2020benchmark,sun2020energy,luo2020accelerating,li20203d,imani2019floatpim,cheng2018time,marinella2018multiscale,hasan2017fast}. In contrast, executing DL training using DRAM-based PIM architectures is challenging, since the area and power constraints of such architectures lead to performance bottlenecks when executing key operations (e.g., multiplication) required during training~\cite{oliveira2021.SLS}. 

\textbf{PIM for other ML algorithms.} Few related prior works~\cite{falahati2018origami, 8645905,sun2020one,shelor2019reconfigurable,gao.pact15,saikia2019k} propose solutions for ML algorithms other than DL inference and training (e.g.,  regression, classification, clustering). Such works leverage different memory technologies (e.g., 3D-stacked DRAM~\cite{falahati2018origami,shelor2019reconfigurable,gao.pact15}, ReRAM~\cite{sun2020one}, SRAM~\cite{8645905,saikia2019k}) to accelerate  ML workloads such as linear regression~\cite{falahati2018origami,8645905,sun2020one,shelor2019reconfigurable}, logistic regression~\cite{falahati2018origami,sun2020one}, support vector \juang{machines}~\cite{falahati2018origami}, and K-nearest neighbors~\cite{8645905,saikia2019k}. None of these works provide comprehensive implementation and evaluation  of ML algorithms using a \emph{real} processing-in-memory architecture.

\section{Conclusion}

Machine learning training frequently becomes memory-bound in processor-centric systems due to repeated accesses to large training datasets. 
Memory-centric systems (i.e., \juang{systems with processing-in-memory (PIM)} capabilities) can overcome this memory boundedness. 

We implement several representative classic machine learning algorithms on a real-world general-purpose PIM architecture with the aim of understanding the potential of memory-centric systems for ML training. 
We evaluate our PIM implementations on a memory-centric computing system with more than 2500 PIM cores in terms of accuracy, performance, and scaling characteristics, and compare to state-of-the-art implementations for CPU and GPU. 

To our knowledge, our work is the first one to evaluate training of machine learning algorithms on a real-world PIM architecture. 
\juang{We show that PIM systems can greatly outperform CPUs and GPUs for memory-bound ML training workloads when the PIM processing elements have native support for the arithmetic operations and datatypes required by the ML training workloads. 
Compared to CPUs, PIM systems feature significantly higher memory bandwidth and many more parallel processing elements, the number of which scales with memory capacity. 
Compared to GPUs, PIM systems benefit from higher host-accelerator bandwidth given that PIM processing elements are connected to the host CPU via memory channels (as opposed to PCIe like GPUs). 
We believe that our work shows great promise for PIM systems as widely-used accelerators for ML training workloads, and this promise can materialize in future PIM systems with more mature architectures, hardware, and software support.}

\ignore{
\subsection{Future Work}
\sylvan{The initial centroids in the \km algorithm in this work are chosen at random. A more efficient way to select them is the \emph{\km++} method~\cite{vassilvitskii2006k}. This initialization method could be implemented on PIM to accelerate convergence.}

\sylvan{The implemented version of the \km algorithm is the original from Lloyd's publication~\cite{Lloyd82leastsquares}. There exists other versions, such as the Elkan variation~\cite{elkan2003using} which is more efficient but more memory-intensive, thus a good candidate for a PIM implementation.}

\sylvan{In order to take full advantage of the PIM architecture scaling on small to medium sized datasets, a greater degree of high-level parallelism would need to be implemented.
\begin{enumerate}
\item For KME, multiple clustering runs with different random initializations could be executed in parallel, with each thread using only a fraction of the available DPUs.
\item For classification, multiple decision trees could be built in parallel to generate a random forest classifier.
\end{enumerate}
Such job level parallelism will be the subject of future works.
}
}

\begin{acks}
We acknowledge the generous gifts provided by our industrial partners, including ASML, Facebook, Google, Huawei, Intel, Microsoft, and VMware.
We acknowledge support from the Semiconductor Research Corporation, the ETH Future Computing Laboratory, \juanggr{and the European Union's Horizon programme for research and innovation under grant agreement No. 101047160, project BioPIM (Processing-in-memory architectures and programming libraries for bioinformatics algorithms)}.
This research was partially supported by ACCESS – AI Chip Center for Emerging Smart Systems, sponsored by InnoHK funding, Hong Kong SAR.

A much shorter version of this paper appears as an invited paper at the 2022 IEEE Computer Society Annual Symposium on VLSI (ISVLSI).

\end{acks}

\balance
{
  \bstctlcite{IEEEexample:BSTcontrol} 
   \let\OLDthebibliography\thebibliography
  \renewcommand\thebibliography[1]{
    \OLDthebibliography{#1}
    \setlength{\parskip}{0pt}
    \setlength{\itemsep}{0pt}
  }
  \bibliographystyle{IEEEtran}
  \bibliography{references}
}


\end{document}